\documentclass[a4paper]{ar-1col} 
\usepackage{url}

\usepackage{cite}

\usepackage[
colorlinks,citecolor=pastelblue,linkcolor=pastelblue,urlcolor=pastelblue,hyperindex]{hyperref}
\definecolor{pastelblue}{RGB}{20,93,160}
\definecolor{brn}{rgb}{.8,.4,.0}

\usepackage{amssymb}
\usepackage{amsfonts}
\usepackage{epstopdf}
\usepackage{dcolumn}
\usepackage{amsmath}
\usepackage{latexsym,bm}
\usepackage{amsthm}
\usepackage{slashed}
\usepackage{float}
\usepackage{url}
\usepackage{rotating}
\usepackage[normalem]{ulem}
\usepackage{faktor}
\usepackage{tensor}
\usepackage{array}
\usepackage{cleveref}

\usepackage{tabularx}
\usepackage{makecell}

\usepackage{tikz}
\usepackage{tikz-cd}
\usetikzlibrary{arrows}
\usetikzlibrary{intersections}
\usetikzlibrary{shapes.geometric}
\usetikzlibrary{decorations.pathmorphing, patterns,shapes}
\usetikzlibrary{decorations.markings}
\usetikzlibrary{calc}

 \numberwithin{equation}{section}

\newcommand{\be}{\begin{equation}}
\newcommand{\ee}{\end{equation}}
\newcommand{\ba}{\begin{aligned}}
\newcommand{\ea}{\end{aligned}}

\newcommand{\Z}{{\mathbb Z}}
\newcommand{\R}{{\mathbb R}}
\newcommand{\C}{{\mathbb C}}

\newcommand{\SPT}{\text{SPT}}
\newcommand{\SymTFT}[3]{\Big( #1 \,\Big|\, #2 \,\Big|\, #3 \Big) }

\newcommand{\cA}{\mathcal{A}}
\newcommand{\cB}{\mathcal{B}}

\newcommand{\cD}{\mathcal{D}}

\newcommand{\cL}{\mathcal{L}}
\newcommand{\cO}{\mathcal{O}}
\newcommand{\cS}{\mathcal{S}}
\newcommand{\cT}{\mathcal{T}}
\newcommand{\cZ}{\mathcal{Z}}
\newcommand{\cP}{\mathcal{P}}
\newcommand{\cI}{\mathcal{I}}
\newcommand{\cN}{\mathcal{N}}

\newcommand{\fB}{\mathfrak{B}}
\newcommand{\fZ}{\mathfrak{Z}}

\newcommand{\Symtft}{\text{SymTFT}}

\newcommand{\Dir}{\text{Dir}}
\newcommand{\Bsym}{\mathfrak{B}^{\text{sym}}}
\newcommand{\Bphys}{\mathfrak{B}^{\text{phys}}}
\newcommand{\Hom}{\text{Hom}}

\renewcommand{\Vec}{\mathsf{Vec}}
\newcommand{\Rep}{\mathsf{Rep}}

\newcommand{\Ising}{\mathsf{Ising}}

\newcommand{\IR}{\text{IR}}

\usepackage{mathrsfs}  
\usepackage{physics}

\newcommand{\Uone}{\mathrm{U}(1)}
\newcommand{\eps}{\epsilon}
\newcommand{\ga}{\gamma}
\newcommand{\om}{\omega}

\newcommand{\si}{\sigma}
\newcommand{\pp}{\partial}

\newcommand{\ii}{\hspace{1pt}\mathrm{i}\hspace{1pt}}
\newcommand{\e}{\hspace{1pt}\mathrm{e}}
\renewcommand{\dd}{\hspace{1pt}\mathrm{d}}
\newcommand{\Lk}{\mathrm{Lk}}

\newcommand{\scrL}{\mathscr{L}}
\newcommand{\scrH}{\mathscr{H}}

\newcommand{\cH}{\mathcal{H}}
\newcommand{\cQ}{\mathcal{Q}}
\newcommand{\TV}{\mathrm{TV}}

\def\wdg{{\mathchoice{\,{\scriptstyle\wedge}\,}{{\scriptstyle\wedge}}{{\scriptscriptstyle\wedge}}{{\scriptscriptstyle\wedge}}}} 

\definecolor{SymTFTcolor}{HTML}{CBD5E8}
\definecolor{PauliColor}{HTML}{B1002C}

\makeatletter 
\AddToHook{begindocument/end}{%
  \def\tagform@#1{\maketag@@@{(\ignorespaces#1\unskip\@@italiccorr)}}%
  \@ifpackageloaded{natbib}{%
    \renewcommand\NAT@open{[}%
    \renewcommand\NAT@close{]}%
  }{%
    \def\@cite#1#2{[{#1\if@tempswa , #2\fi}]}%
  }%
}
\makeatother

\jname{Submitted to Annual Review of Condensed Matter Physics}
\jyear{2026}

\begin{document}

\markboth
  {Pace\kern3.5pt\arblcirc\kern3.5pt
   Sch{\"a}fer-Nameki\kern3.5pt\arblcirc\kern3.5pt Zhang}
  {SymTFT and Phases of Matter}

\title{The SymTFT and Quantum Phases of Matter}

\author{Salvatore D. Pace,$^\alpha$ Sakura Sch{\"a}fer-Nameki,$^\beta$
and Carolyn Zhang$^\gamma$
\affil{$^\alpha$School of Natural Sciences, Institute for Advanced Study, Princeton, NJ 08540, USA}
\affil{$^\beta$Mathematical Institute, University of Oxford, 
Andrew Wiles Building, Woodstock Road, Oxford OX2 6GG, UK}
\affil{$^\gamma$Department of Physics and Astronomy and Quantum Matter Institute, University of British Columbia, Vancouver, BC V6T 1Z1, Canada}
}

\begin{abstract}
The symmetry topological field theory (SymTFT) provides a unified framework for ordinary and generalized symmetries of quantum many-body systems. 
By encoding symmetry in a topological theory in one higher dimension, it separates symmetry data from the details of a specific quantum system. 
In this review, we give a pedagogical introduction to this framework and survey its applications to phases of quantum matter. 
For fusion category symmetries in ${(1+1)}$d, we explain how the SymTFT characterizes anomalies, classifies gapped phases, and organizes the symmetry structure of gapless phases. 
We also give a broad overview of further developments of this framework, including extensions to symmetries in higher dimensions, continuous symmetries, fermionic systems, mixed states, and spacetime symmetries.
\\ 
\end{abstract}

\maketitle

\tableofcontents

\section{Introduction}

Symmetries are fundamental in organizing our description of physical phenomena. 
Much of the structure that symmetries impose, along with their consequences, can be established irrespective of the particular {Hamiltonian or state} describing a system.
For example, the representation theory of a symmetry group $G$ organizes states and operators into multiplets, also called symmetry charges, constrains the allowed interactions, and helps characterize phases of matter.
Following the identification of symmetries with topological defects in~\cite{GKS14125148}, the concept of symmetry has been substantially extended and generalized (see Refs.~\cite{
M220403045,S230518296,
Bhardwaj:2023kri,
Luo:2023ive, 
S230800747} for reviews). 
Many physical systems, from the critical Ising model to the Standard Model of particle physics, have these generalized symmetries, which provide new tools, constraints, and organizing principles for understanding their behavior.

Non-invertible symmetries generalize conventional symmetries by including topological defects without an inverse under fusion.
They go beyond group theory since every element of a group has an inverse.
Topological defects for conventional symmetries fuse according to the group law: ${D_g\otimes D_h = D_{gh}}$. For non-invertible symmetries, the fusion rules instead take the form
\be\label{NonInvFusion}
D_i \otimes D_j = \bigoplus_{k} N_{ij}^k \,D_k \,,\qquad  N_{ij}^k \in \Z_{\geq 0} \,.
\ee
The direct sum implies that ${\langle  \cdots \left(\bigoplus_{k} N_{ij}^k \, D_k \right) \cdots \rangle = \sum_{k} N_{ij}^k\, \langle \cdots D_k \cdots  \rangle}$.
For example, the Kramers--Wannier symmetry defect $D_\si$ of the critical Ising model is non-invertible and satisfies ${D_\si \otimes D^\dagger_\si = 1\oplus D_\psi}$, where $D_\psi$ generates the $\Z_2$ spin-flip symmetry.

\begin{marginnote}[]
\entry{Topological defect}{}
In this review, we call a defect a topological defect if its support can be smoothly deformed without changing correlation functions, provided no other insertions are crossed. This meaning of ``topological'' differs from that used for vortices and hedgehogs, which are classified by the topology of an order-parameter space.
\end{marginnote}

Taking full advantage of generalized symmetries requires mathematical tools that extend the role of representation theory in labeling symmetry charges and characterizing phases. 
The corresponding generalized charges, the analogues of group representations, require a richer mathematical description because non-invertible symmetries can relate genuine local operators to operators attached to topological defects.
The mathematics underlying these tools is category theory, whose development remains an active area of mathematical research, whereas much of the group theory used in physics was established over a century ago.

The symmetry topological field theory (SymTFT) has emerged as a framework for systematically applying these categorical tools to understand the consequences of generalized symmetries~\cite{
GKS14125148,
JW191213492,
KLW200514178,
GK200805960, 
ABG211202092, 
FT220907471, 
Bhardwaj:2023ayw}.
It is sometimes called topological holography (although it does not include gravity)~\cite{Moradi:2022lqp}. 
In this framework, the symmetry of a quantum many-body system or quantum field theory is encoded in a topological field theory (TFT) in one higher dimension---the SymTFT---with two boundary conditions. 
The SymTFT may be invertible or have intrinsic topological order, and its topological order is sometimes called the symmetry topological order (SymTO). 
We consider the SymTFT on a space with two boundaries.
Its first boundary condition specifies the symmetry, and is known as the symmetry boundary. 
The second boundary, known as the physical boundary, specifies the particular quantum theory. This structure allows the symmetry to be studied independently of a specific Hamiltonian or Lagrangian.
The topological defects, e.g., the anyons, of the SymTFT encode the generalized charges of a symmetry, and the boundary conditions provide a framework for classifying gapped phases and organizing the symmetry structure of gapless theories.

This review provides an accessible introduction to the SymTFT and surveys its applications to phases of quantum matter with generalized symmetry.
We primarily focus on ${(1+1)}$d systems whose symmetries are described by fusion categories.
In Sec.~\ref{Pedagogicalintro}, we introduce the central concepts of the SymTFT through the familiar setting of abelian group symmetries. 
We then extend this to non-invertible symmetries in Sec.~\ref{Sec:FusionCatSymTFT}, and survey the SymTFT's applications to generalized charges, gapped and gapless phases, and anomalies. 
We discuss generalizations to fusion 2-category symmetries in ${(2+1)}$d in Sec.~\ref{sec:Gapped2plus1d}.
Perhaps the most powerful application of the SymTFT we discuss is the classification of gapped phases \cite{KLW200514178,Garre-Rubio:2022uum,Bhardwaj:2023ayw, BBP231003784}, which we discuss in Secs.~\ref{GroupSectionGappedPhases} and~\ref{fusionCatGappedPhaseSec} for groups and fusion category symmetries, respectively, and in Sec.~\ref{sec:Gapped2plus1d} for fusion 2-category symmetries. 
Finally, we conclude in Sec.~\ref{sec:outlook} with a summary of other generalizations and applications of the SymTFT, and discuss some open problems.

\section{What is the SymTFT? A Pedagogical Introduction.}\label{Pedagogicalintro}

This section introduces the SymTFT paradigm and its central concepts through the familiar setting of finite abelian group symmetries in ${(1+1)}$d quantum systems.

\subsection{The SymTFT Paradigm}\label{sec:symtftParadigm}

At its most basic level, the SymTFT provides a useful way of organizing the data of a quantum theory. 
In this paradigm, a quantum theory $\mathcal{T}$ on a $(d+1)$-dimensional spacetime $M$ admits an equivalent description in terms of a $(d+2)$-dimensional topological theory on $M \times I$, where $I$ is the unit interval, equipped with suitable boundary conditions. 
Its defining purpose is to separate the \emph{symmetry} data $\mathcal{S}$ of $\mathcal{T}$ from the \emph{theory-specific} details of $\mathcal{T}$, such as its specific degrees of freedom and Hamiltonian/Lagrangian.

The bulk ${(d+2)}$-dimensional topological theory is a topological field theory (TFT)
determined solely by the symmetry $\cS$, and is called the symmetry TFT (SymTFT) of $\cS$. We denote it by $\fZ(\cS)$. 
The boundary $\Bsym$ on ${M\times \{0\}}$ is called the symmetry boundary and it encodes the symmetry $\cS$.
Both $\fZ(\cS)$ and $\Bsym_\cS$ depend only on $\cS$, and not on the theory-specific details of $\cT$. 
The other boundary, ${\Bphys_\cT}$ at ${M\times \{1\}}$, is called the physical boundary and it encodes the theory-specific data of $\cT$. 
We denote the equivalence between $\cT$ and this topological theory in one higher dimension by
\begin{equation}\label{SymtftDecomposition}
    \cT = \SymTFT{\Bsym_\cS}{\fZ (\cS)}{\Bphys_\cT} \,.
\end{equation}

The theory ${(\Bsym_\cS \mid \fZ(\cS) \mid \Bphys_\cT)}$ is often called the \emph{SymTFT sandwich} presentation of $\cT$. 
This presentation of $\cT$ is useful because it separates questions about symmetry from questions about the particular dynamics of the theory. For instance, kinematical questions involving $\cS$, such as gauging and anomalies, can be addressed by studying the symmetry boundary $\Bsym_\cS$, while dynamical questions about $\cT$, such as its possible phases, are answered using the physical boundary $\Bphys_\cT$.

\begin{marginnote}[]
\entry{SymTFT Sandwich}{
\begin{equation*}
\begin{tikzpicture}
\draw [SymTFTcolor,  fill=SymTFTcolor] 
(0,0) -- (0,2) -- (1.801,2) -- (1.801,0) -- (0,0) ; 
\draw [very thick] (0,0) -- (0,2) ;
\draw [very thick] (1.801,0) -- (1.801,2) ;
\node at  (1.801/2,1)  {$\fZ(\cS)$} ;
\node[above] at (0,2) {$\Bsym_{\cS}$}; 
\node[above] at (1.801,2) {$\Bphys_{\cT}$}; 
\end{tikzpicture}
\end{equation*}
}
\end{marginnote}

The decomposition~\eqref{SymtftDecomposition} applies to ordinary and generalized symmetries $\cS$, including non-invertible, higher-form, and subsystem symmetries.
We discuss the SymTFT for categorical symmetries in Sec.~\ref{Sec:FusionCatSymTFT}. 
In this section, we specialize to the case where $\cS$ is a finite group $G$, for which $\fZ(G)$ is $G$ gauge theory in ${d+2}$ spacetime dimensions. The simplest non-trivial example is an anomaly-free ${G=\Z_2}$ symmetry. 
We now illustrate~\eqref{SymtftDecomposition} explicitly in this simple case from two complementary perspectives: the continuum field theory perspective and the quantum lattice system perspective. Readers more familiar with lattice models may begin with Sec.~\ref{sec:Z2SymLatEx}; however, both examples realize the same construction.

\subsubsection{Example: $\Z_2$ Symmetry (Continuum Perspective)}\label{sec:Z2SymContEx}

Consider a ${(1+1)}$d quantum field theory on Euclidean spacetime $M_2$ with a $\Z_2$ $0$-form global symmetry. For concreteness, we assume the theory has a single complex scalar field $\phi$ and Lagrangian
\begin{equation}\label{ScalarFieldTheoryLagrangian}
    \scrL[\phi] = \frac12  |\pp_\mu\phi|^2 + V(|\phi|).
\end{equation}
The $\Z_2$ $0$-form global symmetry transforms ${\phi\mapsto -\phi}$. ($\scrL[\phi]$ has a larger U(1) symmetry, but we focus on this $\Z_2$ subgroup.)
We now construct its SymTFT realization.

The SymTFT of a $\Z_2$ $0$-form global symmetry in ${(1+1)}$d is $\Z_2$ gauge theory in ${(2+1)}$d.
For ${(2+1)}$d Euclidean spacetime $M_3$, its continuum field theory action is
\begin{equation}
    S_{\Symtft}^{\Z_2}[M_3] = \frac{2\ii}{2\pi} \int_{M_3} b \wdg \dd a,
\end{equation}
where $a$ and $b$ are $\Uone$ gauge fields. 
Denoting by $C$ a loop in $M_3$, the topological defect lines of BF theory are $L(q_e, q_m; C) = W(C)^{q_e}\, V(C)^{q_m}$ where ${W(C) = \mathrm{exp}(\ii\oint _C a)}$ and ${V(C) = \mathrm{exp}(\ii\oint_C b)}$.
The defects $W$ and $V$ are \textit{topological} defects because their correlation functions satisfy ${\langle W(C) \rangle = \langle W(C+ \pp D) \rangle}$ and ${\langle V(C) \rangle = \langle V(C+ \pp D) \rangle}$.
They also satisfy ${\langle W^2(C)\rangle  = \langle V^2(C)\rangle = 1}$ and ${\langle W(C_1) \,V(C_2) \rangle = (-1)^{\Lk(C_1,C_2)}}$~\cite{KS14010740}.

To realize the field theory~\eqref{ScalarFieldTheoryLagrangian} using this SymTFT, we take ${(2+1)}$d spacetime ${M_3}$ to be ${M_2\times I}$ with ${I= [0,1]}$ the unit interval. 
The symmetry and physical boundaries are ${\Bsym}$ at ${M_2\times\{0\}}$ and ${\Bphys}$ at ${M_2\times\{1\}}$, respectively.
On the symmetry boundary, we enforce the Dirichlet boundary condition ${a\,|_{\Bsym} = 0}$, which allows the $W$ defect to end on $\Bsym$.
A transformation ${a\mapsto a+\dd\alpha}$ must satisfy ${\dd\alpha\,\big|_{\Bsym} = 0}$, which implies ${\alpha\,\big|_{\Bsym} = \text{constant}}$.
This is a gauge transformation when ${\mathrm{e}^{\ii\alpha\mid_{\Bsym}} = 1}$.
The action on ${M_3 = M_2\times I}$ is
\begin{equation}
    S[M_2] = S_{\Symtft}^{\Z_2}[M_2\times I]
    +
    \int_{M_2 \times \{1\}} \left(\frac12  |(\pp_\mu - \ii a_\mu)\phi|^2 + V(|\phi|) \right) \dd^2 \bm{x}.
\end{equation}
The $\Bphys$ boundary term is the original Lagrangian~\eqref{ScalarFieldTheoryLagrangian} coupled to the gauge field $a$. 
The gauge transformation on $b$ is now ${b\mapsto b+\dd\beta}$ with ${\beta\,\big|_{\Bphys} = 0}$.
Integrating out $b$ causes $a$ to be a $\Z_2$ gauge field.
The field $\phi$ is no longer gauge invariant, but it can be made gauge invariant by attaching it to a $W$ defect line that ends on $\Bsym$: ${\hat{\phi}(x^{\mathrm{phys}}) = \phi(x^{\mathrm{phys}})
\,
\mathrm{exp}(\ii\int_{x^{\mathrm{phys}}}^{x^{\mathrm{sym}}}a) }$.
Under ${(a,\phi)\mapsto (a+\dd\alpha,\mathrm{e}^{\ii\alpha}\phi )}$, this transforms as ${\hat{\phi}\mapsto \mathrm{e}^{\ii\alpha\mid_{\Bsym}}\,\hat{\phi}}$. Hence, $\hat{\phi}$ is gauge-invariant, and the $\Z_2$ symmetry transformation of~\eqref{ScalarFieldTheoryLagrangian} is ${(a,\phi)\mapsto (a+\dd\alpha,\mathrm{e}^{\ii\alpha}\phi )}$ with ${\alpha\,\big|_{\Bsym} = \pi\pmod{2\pi}}$.
It follows from ${\langle W(C_1) \,V(C_2) \rangle = (-1)^{\Lk(C_1,C_2)}}$ that the $\Z_2$ symmetry defect is the defect line $V$ on $\Bsym$. Importantly, in line with the general SymTFT paradigm, the $\Z_2$ symmetry defect depends only on the fields of the bulk SymTFT and not on the $\Bphys$ theory.

\subsubsection{Example: $\Z_2$ Symmetry (Lattice Perspective)}\label{sec:Z2SymLatEx}

Consider a ${(1+1)}$d quantum lattice system with an onsite $\Z_2$ symmetry. For simplicity, we have a single qubit residing on each site ${j}$ of the lattice, and denote its Pauli operators by $X_{j}$ and $Z_{j}$. 
We further assume the lattice has $L$ sites with periodic boundary conditions ${j\sim j+L}$.
Suppose the $\Z_2$ symmetry is generated by the unitary operator ${\eta = \prod_{j=1}^L X_{j}}$.
A simple family of Hamiltonians that commute with $\eta$ is the transverse-field Ising model
\begin{equation}
    H(h) = -\sum_{j=1}^L (Z_{j} Z_{j+1} + h\, X_{j}).
\end{equation}

The SymTFT of an onsite $\Z_2$ symmetry can be realized by the ground state space of the toric code Hamiltonian. (Equivalently, by the toric code with the bulk gap sent to infinity.)
Consider, for simplicity, a square lattice of ${L_x\times L_y}$ sites with a single qubit on each edge and periodic boundary conditions. We denote a generic vertex by ${v\equiv (x,y)}$ and the edge connecting $v$ and ${v+\hat{\mu}}$, with ${\mu=x,y}$, by $v,\mu$.
The toric code Hamiltonian is ${ - \sum_v (A_v + B_v)}$, where the mutually commuting stabilizers are
\begin{equation}\label{TC stabs}
    A_v =
    \begin{tikzpicture}[scale = 0.4, baseline = {([yshift=-.5ex]current bounding box.center)}]
        \coordinate (L) at (0,0);
        \coordinate (R) at (6*7/10,0);
        \coordinate (B) at (3*7/10,-3*7/10);
        \coordinate (T) at (3*7/10,3*7/10);
        \coordinate (mid) at (3*7/10,0);
        \coordinate (midL) at (1.5*7/10,0);
        \coordinate (midR) at (4.5*7/10,0);
        \coordinate (midB) at (3*7/10,-1.5*7/10);
        \coordinate (midT) at (3*7/10,1.5*7/10);
        \draw[line width=0.015in, black] (L) -- (R);
        \draw[line width=0.015in, black] (B) -- (T);
        \fill[white] (midL) circle (10pt);
        \node[PauliColor] at (midL) {\normalsize $X$};
        \fill[white] (midR) circle (10pt);
        \node[PauliColor] at (midR) {\normalsize $X$};
        \fill[white] (midB) circle (12pt);
        \node[PauliColor] at (midB) {\normalsize $X$};
        \fill[white] (midT) circle (12pt);
        \node[PauliColor] at (midT) {\normalsize $X$};
        \fill[black] (mid) circle (6pt);
    \end{tikzpicture}\,,
    \qquad 
    B_v  = 
    \begin{tikzpicture}[scale = 0.4, baseline = {([yshift=-.5ex]current bounding box.center)}]
        \coordinate (bL) at (0,0);
        \coordinate (bR) at (3,0);
        \coordinate (tL) at (0,3);
        \coordinate (tR) at (3,3);
        \coordinate (b) at (1.5,0);
        \coordinate (t) at (1.5,3);
        \coordinate (l) at (0,1.5);
        \coordinate (r) at (3, 1.5);
        \draw[line width=0.015in, black] (bL) -- (bR) -- (tR) -- (tL) -- cycle;
        \fill[white] (b) circle (10pt);
        \node[PauliColor] at (b) {\normalsize $Z$};
        \fill[white] (r) circle (12pt);
        \node[PauliColor] at (r) {\normalsize $Z$};\fill[white] (t) circle (10pt);
        \node[PauliColor] at (t) {\normalsize $Z$};\fill[white] (l) circle (12pt);
        \node[PauliColor] at (l) {\normalsize $Z$};
        \fill[black] (bL) circle (6pt);
    \end{tikzpicture}.
\end{equation}
The black disk represents the vertex $v$. Thus, the SymTFT Hilbert space is
\begin{equation}
    \scrH_{\Symtft}^{\Z_2} = \big\{ \ket{\psi}\in \bigotimes_e \C^2 \mid A_v \ket{\psi} = B_v \ket{\psi} = \ket{\psi}\text{ for all }v\big\}.
\end{equation}
The operators acting on this Hilbert space are formed by ${W(C) = \prod_{(v,\mu)\subset C} Z_{v,\mu}}$ and ${V(C^\vee) = \prod_{(v,\mu)\subset C^\vee} X_{v,\mu}}$, where $C$ is a loop on the lattice and $C^\vee$ is a loop on the dual lattice. 
$W(C_1)$ and $V(C^\vee_2)$ anti-commute when $C_1$ and $C^\vee_2$ intersect an odd number of times, and otherwise commute.
Since ${A_v = B_v = 1}$ on $\scrH_{\Symtft}^{\Z_2}$, $W$ and $V$ depend only on the homology of $C$ and $C^\vee$.

To realize the transverse-field Ising model using this SymTFT, we consider the SymTFT on a spatial cylinder with periodic boundary conditions in the $y$-direction and spatial boundaries $\Bsym$ and $\Bphys$ at ${x=1}$ and ${x = L_x}$, respectively. A single qubit resides on each edge except vertical edges on $\Bsym$. Define
\begin{equation}
    B^{(\text{rough})}_v  =~
    \begin{tikzpicture}[scale = 0.4, baseline = {([yshift=-.5ex]current bounding box.center)}]
        \coordinate (bL) at (0,0);
        \coordinate (bR) at (3*7/8,0);
        \coordinate (tL) at (0,3);
        \coordinate (tR) at (3*7/8,3);
        \coordinate (b) at (1.5*7/8,0);
        \coordinate (t) at (1.5*7/8,3);
        \coordinate (l) at (0,1.5);
        \coordinate (r) at (3*7/8, 1.5);
        \draw[line width=0.015in, black] (bL) -- (bR);
        \draw[line width=0.015in, black] (bR) -- (tR);
        \draw[line width=0.015in, black] (tL) -- (tR);
        \fill[white] (b) circle (10pt);
        \node[PauliColor] at (b) {\normalsize $Z$};
        \fill[white] (r) circle (12pt);
        \node[PauliColor] at (r) {\normalsize $Z$};\fill[white] (t) circle (10pt);
        \node[PauliColor] at (t) {\normalsize $Z$};
        \fill[black] (bL) circle (6pt);
    \end{tikzpicture},
    \qquad\qquad
    \hat{Z}_j = \prod_{x'=1}^{L_x-1} Z_{(x',j),x},
    \qquad\qquad
    \hat{X}_j = \begin{tikzpicture}[scale = 0.4, baseline = {([yshift=-.5ex]current bounding box.center)}]
        \coordinate (L) at (0,0);
        \coordinate (R) at (6*7/10,0);
        \coordinate (B) at (3*7/10,-3*7/10);
        \coordinate (T) at (3*7/10,3*7/10);
        \coordinate (mid) at (3*7/10,0);
        \coordinate (midL) at (1.5*7/10,0);
        \coordinate (midR) at (4.5*7/10,0);
        \coordinate (midB) at (3*7/10,-1.5*7/10);
        \coordinate (midT) at (3*7/10,1.5*7/10);
        \draw[line width=0.015in, black] (L) -- (mid);
        \draw[line width=0.015in, black] (B) -- (T);
        \fill[white] (midL) circle (10pt);
        \node[PauliColor] at (midL) {\normalsize $X$};
        \fill[white] (midB) circle (12pt);
        \node[PauliColor] at (midB) {\normalsize $X$};
        \fill[white] (midT) circle (12pt);
        \node[PauliColor] at (midT) {\normalsize $X$};
        \fill[black] (mid) circle (6pt);
    \end{tikzpicture}
    ,
\end{equation}
with $B^{(\text{rough})}_v$ belonging to ${v\in \Bsym}$, $\hat{Z}_j$ stretching from $\Bsym$ to $\Bphys$, and $\hat{X}_j$ belonging to ${(L_x,j)\in \Bphys}$.
We choose the $\Bsym$ boundary conditions to be the rough boundary of the toric code, which has $B^{(\text{rough})}_v$ as its boundary stabilizers ~\cite{BK9811052}.
This boundary condition trivializes the $W(C)$ loop operator when $C$ is a non-contractible cycle in the $y$-direction (i.e., the $W$ line can end on the rough boundary).
 The Hilbert space is the eigenspace $\ket{\psi}$ satisfying ${A_v \ket{\psi} = B_v \ket{\psi} = \ket{\psi}}$ for ${v\not\in\Bsym \cup \Bphys}$ and ${B^{(\text{rough})}_v \ket{\psi} = \ket{\psi}}$ for ${v\in\Bsym}$.
The operators $\hat{X}_j$ and $\hat{Z}_j$ preserve these constraints and act as $L_y$ independent Pauli pairs, generating the operator algebra of the Hilbert space.
This space has dimension $2^{L_y}$ and is isomorphic to the transverse-field Ising model's Hilbert space with ${L_y = L}$. On this space, the onsite $\Z_2$ symmetry operator is ${\eta = \prod_{j=1}^{L_y} \hat{X}_j = V(C^\vee)}$ with $C^\vee$ any non-contractible loop.
The transverse-field Ising model Hamiltonian is
\begin{equation}
    -\sum_{j=1}^{L_y} (\hat{Z}_{j} \hat{Z}_{j+1} + h\, \hat{X}_{j})
    =
    -\sum_{v\in \Bphys} (Z_{v,y} + h\, 
    X_{v-\hat{x},x}
    X_{v-\hat{y},y}
    X_{v,y}
).
\end{equation}
Note that when ${h=0}$ (resp. ${h\to\infty}$), $\Bphys$ becomes the rough (resp. smooth) boundary in the ground state subspace.

\subsection{The Symmetry Boundary, Anyons, and Gauging}

As mentioned, the SymTFT $\fZ(\cS)$ associated with a symmetry $\cS$ admits a gapped boundary condition $\Bsym_\cS$ that realizes $\cS$. The pair $(\fZ(\cS),\Bsym_\cS)$ encodes the complete data of the symmetry. 
To study $\cS$ itself, one often omits the physical boundary and considers ${(\Bsym_\cS\mid \fZ(\cS)}$, which is called the \emph{SymTFT quiche} of $\cS$~\cite{FT220907471}.

\begin{marginnote}[]
\entry{SymTFT Quiche}{
$$
\begin{tikzpicture}
\draw [SymTFTcolor,  fill=SymTFTcolor] 
(0,0) -- (0,2) -- (2,2) -- (2,0) -- (0,0) ; 
\draw [white] (0,0) -- (0,2) -- (2,2) -- (2,0) -- (0,0)  ; 
\draw [very thick] (0,0) -- (0,2) ;
\node at  (1,1)  {$\fZ(\cS)$} ;
\node[above] at (0,2) {$\Bsym_{\cS}$}; 
\end{tikzpicture}
$$
}
\end{marginnote}

In the abelian examples considered here, the symmetry boundary $\Bsym_{\cS}$ is characterized by a maximal collection of anyons in $\fZ(\cS)$ that can end topologically on $\Bsym_{\cS}$. Here, by anyons we mean the topological defect/operator lines of $\fZ(\cS)$, rather than gapped quasiparticle excitations.  
Also, an anyon ending (topologically) on the boundary means it ends on a genuine local (topological) operator on the boundary. 
(Recall that a genuine local operator can be inserted independently at a point, whereas a non-genuine local operator is defined only as the endpoint of an attached extended defect, such as a topological line.)
If an anyon ends on a non-genuine local operator on the boundary, we say that anyon turns into a boundary defect.
The anyons that can end topologically on $\Bsym_{\cS}$ encode the symmetry charges of $\cS$. By contrast, anyons that cannot end on $\Bsym_{\cS}$ can be topologically deformed to the boundary, where they become topological line defects/operators realizing the symmetry $\cS$. The relation between these bulk anyons and boundary defects is encoded by the bulk-to-boundary map.

\noindent{\bf Example: $\Z_2$ symmetry.}
The $\Z_2$ symmetry example discussed in Sec.~\ref{sec:symtftParadigm} can be formulated directly in terms of anyons. 
Recall that ${(2+1)}$d $\Z_2$ gauge theory has four anyon types, $1$, $e$, $m$, and $f$, where $1$ denotes the trivial anyon.
The anyons $e$ and $m$ are bosons, whereas $f$ is a fermion. 
Their fusion rules include ${e\otimes e = m\otimes m = 1}$ and  ${f = e\otimes m}$.
The $e$ and $m$ anyons are mutual semions.
The topological line defects/operators from Secs.~\ref{sec:Z2SymContEx} and~\ref{sec:Z2SymLatEx} are identified with these anyons: $W$ corresponds to the electric anyon $e$, while $V$ corresponds to the magnetic anyon $m$. 

\begin{marginnote}[]
\entry{$e$ and $m$ condensed boundaries}{
$$
\begin{tikzpicture}
\begin{scope}[shift={(-2,0)}]
\draw [SymTFTcolor,  fill=SymTFTcolor] 
(0,0) -- (0,2) -- (2,2) -- (2,0) -- (0,0) ; 
\draw [white] (0,0) -- (0,2) -- (2,2) -- (2,0) -- (0,0)  ; 
\draw [very thick] (0,0) -- (0,2) ;
\draw[thick, blue] (0,1.25) -- (2,1.25);
\node[above, blue] at (1,1.25) {$e$};
\draw[thick, red] (0,0.75) -- (2,0.75);
\draw[very thick, red] (0,0.75) -- (0,0);
\node[above, red] at (1,0.75) {$m$};
\node[right, red] at (0,0.75/2) {$m_\text{bdry}$};
\node[above, black] at (0,2) {$\cL_e$}; 
\fill[blue] (0, 1.25) circle (0.05cm);
\fill[red] (0, 0.75) circle (0.05cm);
\node at  (1.5,0.25)  {$\fZ(\Z_2)$};
\end{scope}
\begin{scope}[shift={(-2,-3)}]
    \draw [SymTFTcolor,  fill=SymTFTcolor] 
    (0,0) -- (0,2) -- (2,2) -- (2,0) -- (0,0) ; 
    \draw [very thick] (0,0) -- (0,2) ;
    \draw[thick, blue] (0,1.25) -- (2,1.25);
    \node[above, blue] at (1,1.25) {$e$};
    \draw[thick, red] (0,0.75) -- (2,0.75);
    \draw[very thick, blue] (0,1.25) -- (0,2);
    \node[above, red] at (1,0.75) {$m$};
    \node[right, blue] at (0,1.25+0.75/2) {$e_\text{bdry}$};
    \node[above, black] at (0,2) {$\cL_m$}; 
    \fill[blue] (0, 1.25) circle (0.05cm);
    \fill[red] (0, 0.75) circle (0.05cm);
    \node at  (1,0.25)  {$\fZ(\Z_2)$};
\end{scope}
\end{tikzpicture}
$$
}
\end{marginnote}

There are two gapped boundary conditions of ${(2+1)}$d $\Z_2$ gauge theory specified by which anyons can end. The first is the $e$-condensed boundary ${\cL_e = 1\oplus e}$, on which the anyons $1$ and $e$ can end topologically. This boundary is the Dirichlet boundary of Sec.~\ref{sec:Z2SymContEx} and the rough boundary of Sec.~\ref{sec:Z2SymLatEx}.
The second is the $m$-condensed boundary ${\cL_m = 1\oplus m}$, on which the anyons $1$ and $m$ can end topologically. It is the Neumann boundary of continuum $\Z_2$ gauge theory and the smooth boundary of the toric code.
The bulk-to-boundary map for $\cL_e$, for instance, is ${1,e\xrightarrow{\,\cL_e\,} 1}$ and ${m,f\xrightarrow{\,\cL_e\,} m_{\text{bdry}}}$~\cite{KK11045047}.
The $m$ anyon persists on the $e$-condensed boundary and generates a ${\langle m_\text{bdry}\rangle \cong \Z_2}$ symmetry. Therefore, the $e$-condensed boundary is a $\Z_2$ symmetry boundary.
Similarly, the $m$-condensed boundary is also a $\Z_2$ symmetry boundary: the $e$ anyon persists on it and generates a ${\langle e_\text{bdry}\rangle\cong \Z_2}$ symmetry.

Starting from a SymTFT sandwich
${(\Bsym_{\cS}\mid\fZ(\cS)\mid\Bphys_{\cT})}$, one can gauge the $\cS$ symmetry of $\cT$ by replacing the symmetry boundary $\Bsym_{\cS}$ with another gapped boundary $\Bsym_{\cS^\prime}$ while keeping the physical boundary fixed~\cite{GK200805960}. 
This implements a possibly partial or generalized gauging because the new boundary condenses the bulk lines representing $\cS$ symmetry defects, thereby summing over their defect networks and trivializing (part of) the $\cS$ symmetry~\cite{BT170402330}.
The resulting sandwich ${(\Bsym_{\cS^\prime}\mid\fZ(\cS)\mid\Bphys_{\cT})}$ describes the gauged theory, whose new symmetry $\cS^\prime$ is realized on the new symmetry boundary. Thus, inequivalent gapped boundaries of $\fZ(\cS)$ label inequivalent ways of gauging $\cS$. 
In the $\Z_2$ symmetry example, replacing the $e$-condensed boundary with the $m$-condensed boundary gauges the ${\cS = \langle m_\text{bdry}\rangle \cong \Z_2}$ symmetry and leads to a dual ${\cS^\prime = \langle e_\text{bdry}\rangle \cong \Z_2}$ symmetry. In this case, the symmetry after gauging is the same as the original.
More generally, however, $\cS^\prime$ need not be equivalent to $\cS$, and the gapped boundaries of $\fZ(\cS)$, together with their bulk-to-boundary maps, are considerably more sophisticated than those in this $\Z_2$ example.

\subsection{Gapped Phases and Generalized Charges}\label{GroupSectionGappedPhases}

One of the most powerful applications of the SymTFT $\fZ(\cS)$ is the classification of gapped phases characterized by $\cS$. This includes both spontaneous symmetry-breaking (SSB) phases and symmetry-protected topological (SPT) phases.
The SymTFT provides a systematic prescription for enumerating these phases and for identifying their order parameters, including local, string-like, and more general extended order parameters.

Within the SymTFT paradigm, a gapped phase is described by taking the physical boundary to be gapped. The classification then proceeds as follows:
\begin{enumerate}
    \item Fix a symmetry boundary $\Bsym_{\cS}$ that realizes the symmetry $\cS$,
    \item Classify the gapped physical boundary conditions $\Bphys$ of $\fZ(\cS)$.
\end{enumerate}
The equivalence classes of gapped boundary conditions $\Bphys$ do not depend on the choice of symmetry boundary $\Bsym_{\cS}$.
Thus, gapped phases characterized by symmetry $\cS$ are in one-to-one correspondence with equivalence classes of gapped boundary conditions of the SymTFT $\fZ(\cS)$~\cite{TW191202817, KLW200514178, Moradi:2022lqp, BBP231003784}. 
The choice of symmetry boundary determines how each gapped physical boundary is interpreted as a phase with the corresponding symmetry; for example, whether it is an SSB or SPT phase and, in the former case, which symmetry-breaking pattern it realizes.
Consequently, gauging $\cS$ by replacing
${\Bsym_{\cS}\mapsto\Bsym_{\cS^\prime}}$, while leaving $\Bphys$ unchanged, induces a one-to-one correspondence between gapped phases with symmetry $\cS$ and gapped phases with the symmetry after gauging, $\cS^\prime$.

\begin{marginnote}[]
\entry{Gapped Phases for $\Z_2$}{
$$
\begin{tikzpicture}
\begin{scope}[shift={(-2,0)}]
\begin{scope}[shift={(0,0)}]
\draw [SymTFTcolor,  fill=SymTFTcolor] 
(0,0) -- (0,2) -- (1.25,2) -- (1.25,0) -- (0,0) ; 
\draw [white] (0,0) -- (0,2) -- (1.5,2) -- (1.5,0) -- (0,0)  ; 
\draw [very thick] (0,0) -- (0,2) ;
\draw [very thick] (1.25,0) -- (1.25,2) ;
\draw [gray, thick] (0,1) -- (1.25,1);
\draw [blue, thick] (0,1.5) -- (1.25,1.5);
\node at  (1.25/2,0.5)  {$\fZ(\Z_2)$} ;
\node[gray, above] at (1.25/2,1) {$1$};
\node[blue, above] at (1.25/2,1.5) {$e$};
\node[above] at (0,2) {$\cL_e$}; 
\node[above] at (1.25,2) {$\cL_e$}; 
\fill[gray] (0,1) circle (0.05cm);
\fill[gray] (1.25,1) circle (0.05cm);
\fill[blue] (0,1.5) circle (0.05cm);
\fill[blue] (1.25,1.5) circle (0.05cm);
\end{scope}
\begin{scope}[shift={(2,0)}]
\node at (-0.4,1) {$=$};
\draw [very thick] (0,0) -- (0,2) ;
\fill[gray] (0,1) circle (0.05cm);
\fill[blue] (0,1.5) circle (0.05cm);
\node[right] at (0,1) {$\cO_1$};
\node[right] at (0,1.5) {$\cO_e$};
\end{scope}
\begin{scope}[shift={(0,-3)}]
\draw [SymTFTcolor,  fill=SymTFTcolor] 
(0,0) -- (0,2) -- (1.25,2) -- (1.25,0) -- (0,0) ; 
\draw [white] (0,0) -- (0,2) -- (1.25,2) -- (1.25,0) -- (0,0)  ; 
\draw [very thick] (0,0) -- (0,2) ;
\draw [very thick] (1.25,0) -- (1.25,2) ;
\draw [gray, thick] (0,1) -- (1.25,1);
\node at  (1.25/2,0.5)  {$\fZ(\Z_2)$} ;
\node[gray, above] at (1.25/2,1) {$1$};
\node[above] at (0,2) {$\cL_e$}; 
\node[above] at (1.25,2) {$\cL_m$}; 
\fill[gray] (0,1) circle (0.05cm);
\fill[gray] (1.25,1) circle (0.05cm);
\end{scope}
\begin{scope}[shift={(2,-3)}]
\node at (-0.4,1) {$=$};
\draw [very thick] (0,0) -- (0,2) ;
\fill[gray] (0,1) circle (0.05cm);
\node[right] at (0,1) {$\cO_1$};
\end{scope}
\end{scope}
\end{tikzpicture}
$$
}
\end{marginnote}

\begin{marginnote}[]
\entry{${\Z_2\times \Z_2}$ SPT string order parameters}{
$$
\begin{tikzpicture}
\begin{scope}[shift={(0,0)}]
\draw [SymTFTcolor,  fill=SymTFTcolor] 
(0,0) -- (0,2) -- (1.25,2) -- (1.25,0) -- (0,0) ; 
\draw [white] (0,0) -- (0,2) -- (1.25,2) -- (1.25,0) -- (0,0)  ; 
\draw [very thick] (0,0) -- (0,2) ;
\draw [very thick] (1.25,0) -- (1.25,2) ;
\draw [red, thick] (0,1) -- (1.25,1);
\draw [red, very thick] (0,1) -- (0,0);
\node at  (1.25/2,0.5)  {$\fZ(\Z_2^2)$} ;
\node[red, above] at (1.25/2,1) {$m_1$};
\node[above] at (0.125,2) {$\cL_e$}; 
\node[above] at (1.25,2) {$\cL_m$}; 
\fill[red] (0,1) circle (0.05cm);
\fill[red] (1.25,1) circle (0.05cm);
\end{scope}
\begin{scope}[shift={(2,0)}]
\node at (-0.4,1) {$=$};
\draw [very thick] (0,0) -- (0,2) ;
\fill[red] (0,1) circle (0.05cm);
\node[right] at (0,1) {$\cO_{m_1}$};
\draw [red, very thick] (0,1) -- (0,0);
\end{scope}
\begin{scope}[shift={(0,-3)}]
\draw [SymTFTcolor,  fill=SymTFTcolor] 
(0,0) -- (0,2) -- (1.25,2) -- (1.25,0) -- (0,0) ; 
\draw [white] (0,0) -- (0,2) -- (1.25,2) -- (1.25,0) -- (0,0)  ; 
\draw [very thick] (0,0) -- (0,2) ;
\draw [very thick] (1.25,0) -- (1.25,2) ;
\draw [violet, thick] (0,1) -- (1.25,1);
\draw [red, very thick] (0,1) -- (0,0);
\node at  (1.25/2,0.5)  {$\fZ(\Z_2^2)$} ;
\node[violet, above] at (1.25/2,1) {$m_1e_2$};
\node[above] at (0.125, 2) {$\cL_e$}; 
\node[above] at (1.15,2) {$\cL_\SPT$}; 
\fill[violet] (0,1) circle (0.05cm);
\fill[violet] (1.25,1) circle (0.05cm);
\end{scope}
\begin{scope}[shift={(1.975,-3)}]
\node at (-0.35,1) {$=$};
\draw [very thick] (0,0) -- (0,2) ;
\node[right] at (0,1) {$\cO_{m_{\!1}\!e_2}$};
\draw [red, very thick] (0,1) -- (0,0);
\fill[violet] (0,1) circle (0.05cm);
\end{scope}
\end{tikzpicture}
$$
}
\end{marginnote}

\begin{marginnote}[]
\entry{${\Z_2\times \Z_2}$ SPT interface}{
$$
\begin{tikzpicture}
\begin{scope}[shift={(-2,0)}]
\begin{scope}[shift={(0,0)}]
\draw [SymTFTcolor,  fill=SymTFTcolor] 
(0,0) -- (0,2) -- (2,2) -- (2,0) -- (0,0) ; 
\draw [very thick] (0,0) -- (0,2) ;
\draw [very thick] (2,0) -- (2,2) ;
\draw [blue, thick] (0,1) -- (1,1);
\draw [red, thick] (1,1) -- (2,1.25);
\draw [violet, thick] (1,1) -- (2,0.75);
\node at  (1,0.25)  {$\fZ(\Z_2^2)$} ;
\node[blue, above] at (1/2,1) {$e_2$};
\node[red] at (3/2,1.3) {$m_1$};
\node[violet] at (1.5,0.7) {$e_2m_1$};
\node[above] at (0,2) {$\cL_e$}; 
\node[above] at (2,2) {$\cL_m$}; 
\node[below] at (2,0) {$\cL_{\text{SPT}}$}; 
\fill[blue] (0,1) circle (0.05cm);
\fill[black] (1,1) circle (0.05cm);
\fill[red] (2,1.25) circle (0.05cm);
\fill[violet] (2,0.75) circle (0.05cm);
\filldraw[fill=white, draw=black]
  (2,1) ++(-0.05cm,-0.05cm) rectangle ++(0.1cm,0.1cm);
\end{scope}
\end{scope}
\end{tikzpicture}
$$
}
\end{marginnote}

\noindent{\bf Example: $\Z_2$ symmetry.} To classify ${(1+1)}$d gapped phases with $\Z_2$ symmetry using the SymTFT, we take the bulk theory to be $\fZ(\Z_2)$ and fix the symmetry boundary to condense the anyons in $\cL_e$. Since $\fZ(\Z_2)$ has two inequivalent gapped boundary conditions, $\cL_e$ and $\cL_m$, there are two types of gapped physical boundary and hence two gapped phases~\cite{JW191213492, LTL200304328}:
\begin{equation}
    \SymTFT{\cL_e}{\fZ(\Z_2)}{\cL_e}
    = \Z_2\text{ SSB phase},
    \qquad
    \SymTFT{\cL_e}{\fZ(\Z_2)}{\cL_m}
    = \Z_2\text{ symmetric phase}.
\end{equation}
As illustrated graphically in the margin, the genuine local operators of each sandwich arise from bulk anyons that can end topologically on both the symmetry and physical boundaries. 
In the sandwich ${(\cL_e\mid\fZ(\Z_2)\mid\cL_e)}$, both $1$ and $e$ can end on the two boundaries, giving rise to the local operators $\cO_1$ and $\cO_e$. 
Because $e$ and $m$ are mutual semions, $\cO_e$ is odd under the $\Z_2$ symmetry generated by $m_{\mathrm{bdry}}$.
Since the SymTFT captures renormalization-group (RG) invariant infrared data, the survival of this nontrivial $\Z_2$-odd local operator in the deep infrared (IR) identifies $\cO_e$ as an order parameter and shows that the sandwich describes the $\Z_2$-SSB phase. 
In contrast, only the trivial anyon $1$ can end on both boundaries in the sandwich ${(\cL_e\mid\fZ(\Z_2)\mid\cL_m)}$, so $\cO_1$ is the only genuine local operator. The absence of any nontrivial genuine local operator in the deep IR implies a unique $\Z_2$-invariant ground state, and hence this sandwich describes the $\Z_2$-symmetric phase.

\noindent
{\bf Example: ${\Z_2 \times \Z_2}$ symmetry.} The SymTFT of an anomaly-free ${\Z_2 \times \Z_2}$ symmetry is ${\Z_2 \times \Z_2}$ gauge theory. 
It admits six inequivalent gapped boundary conditions.
We take the symmetry boundary to be the electric boundary ${\cL_e = 1 \oplus e_1 \oplus e_2\oplus e_1e_2}$, which has a ${\Z_2\times\Z_2}$ symmetry generated by $m_{1,\mathrm{bdry}}$ and $m_{2,\mathrm{bdry}}$. Consider two of the gapped boundaries
\begin{equation}
    \cL_m = 1 \oplus m_1 \oplus m_2\oplus m_1m_2,
    \qquad
    \cL_\SPT = 1 \oplus e_1 m_2 \oplus e_2m_1\oplus e_1e_2m_1m_2.
\end{equation}
Both $\cL_m$ and $\cL_{\SPT}$ have no nontrivial anyons in common with $\cL_e$ and, therefore, describe symmetric gapped phases. They are the trivial and nontrivial ${\Z_2\times\Z_2}$ SPT phases, respectively. (The other four gapped boundaries describe SSB phases.)
The distinction between these two SPTs is encoded in their string order parameters. 
For the physical boundary $\cL_m$, the string order parameters are $m_1$ and $m_2$, which become the $m_{1,\mathrm{bdry}}$ and $m_{2,\mathrm{bdry}}$ boundary lines without carrying an additional symmetry charge. 
When $\cL_{\mathrm{SPT}}$ is the physical boundary, the string order parameters are $e_2m_1$ and $e_1m_2$, which turn into boundary defects $m_{1,\mathrm{bdry}}$ and $m_{2,\mathrm{bdry}}$ on $\cL_e$ through junctions carrying $e_2$ and $e_1$ charge, respectively. They, therefore, correspond to non-genuine charged operators of the sandwich theory, which is the decorated domain wall manifestation of the nontrivial ${\Z_2\times\Z_2}$ SPT phase~\cite{CLV13034301}.
The distinction between these SPTs can also be seen using an interface on the physical boundary between $\cL_m$ and $\cL_\SPT$.
The interface supports charged local order parameters, represented in the SymTFT by an electric anyon splitting into two anyons that end on the respective physical boundary segments. Consequently, any symmetry-preserving interface between the trivial and nontrivial SPT phases supports a protected projective zero mode~\cite{WP231100050}.

\noindent{\bf Generalized Charges are Anyons.}
The preceding examples illustrate a general principle of the SymTFT framework: the generalized charges of a symmetry $\cS$ are encoded by the anyons of the SymTFT $\fZ(\cS)$. Their fusion describes how generalized charges combine, while their braiding encodes how they transform under $\cS$. Upon placing the SymTFT on an interval, the bulk-to-boundary maps determine how these charges are realized in the physical theory. An anyon that ends on both boundaries becomes a genuine local operator. An anyon that ends on $\Bphys$ but does not end on $\Bsym$, instead continuing as a nontrivial defect along $\Bsym$, corresponds to a non-genuine local operator attached to that defect.
In this way, the anyon content of $\fZ(\cS)$ provides a unified description of ordinary symmetry charges and generalized order parameters that diagnose the different gapped phases with symmetry $\cS$.

\subsection{General Abelian Symmetry}
\label{sec:Abs}

The preceding examples generalize naturally to any finite abelian group $\mathbb{A}$. We begin with an anomaly-free $\mathbb{A}$ symmetry, whose SymTFT $\fZ(\mathbb{A})$ is the untwisted ${(2+1)}$d $\mathbb{A}$ gauge theory. Its bulk anyons are labeled by pairs ${(a,\chi)\in \mathbb{A}\times\widehat{\mathbb{A}}}$ where $\widehat{\mathbb{A}}\equiv\Hom(\mathbb{A},\Uone)$; 
$a$ labels magnetic flux and $\chi$ electric charge. Their fusion satisfies ${(a,\chi)\otimes(b,\psi)=(a+b,\chi\psi)}$ and, hence, forms the abelian fusion group
$\mathbb{A}\times\widehat{\mathbb{A}}$.
The topological spin of ${(a,\chi)}$ is ${\theta_{(a,\chi)}=\chi(a)}$, implying mutual braiding statistics ${B\bigl((a,\chi),(b,\psi)\bigr)=\chi(b)\psi(a)}$.

Inequivalent gapped boundary conditions of $\fZ(\mathbb{A})$ are classified by Lagrangian subgroups ${L\leq\mathbb{A}\times\widehat{\mathbb{A}}}$~\cite{KS10080654}. 
These are subgroups of order ${|L|=|\mathbb{A}|}$ consisting of only bosonic anyons.
The anyons in $L$ can end topologically on the corresponding gapped boundary, which we specify by ${\cL_L=\bigoplus_{\ell\in L}\ell}$.
As a symmetry boundary, the electric boundary ${\cL_e=\bigoplus_{\chi\in\widehat{\mathbb{A}}}(0,\chi)}$ describes an $\mathbb{A}$ symmetry, whereas the magnetic boundary ${\cL_m=\bigoplus_{a\in\mathbb{A}}(a,1)}$ describes an $\widehat{\mathbb{A}}$ symmetry.

An equivalent and often more physical description labels the gapped boundaries by a subgroup ${H\leq\mathbb{A}}$ and cohomology class ${[\ga]\in \cH^2(H,\Uone)}$.
The pair ${(H,[\ga])}$ can be understood as follows. Starting with the electric boundary, 
one obtains the boundary associated with ${(H,[\ga])}$ by first stacking the boundary with the ${(1+1)}$d $H$-SPT phase ${[\ga]}$ and then gauging $H$. The subgroup $H$ determines which magnetic fluxes are condensed, while $[\ga]$ determines the electric charges attached to those fluxes. The choice ${H=1}$ gives the electric boundary, whereas ${H=\mathbb{A}}$ with ${[\ga]=[1]}$ gives the magnetic boundary. For example, in ${\mathbb{A}=\mathbb{Z}_2\times\mathbb{Z}_2}$ gauge theory discussed above, ${(\mathbb{A},[1])}$ corresponds to $\cL_m$ and ${(\mathbb{A},[\ga_{\mathrm{nt}}])}$ corresponds to ${\cL_{\mathrm{SPT}}}$ where the nontrivial cohomology class attaches charge of one $\mathbb{Z}_2$ symmetry to flux of the other.

An important consequence of this classification is that gauging a subgroup of a non-anomalous symmetry can produce a new symmetry with an anomaly and vice versa (see also Ref.~\cite{tachikawa2017gauging}). The simplest example begins with
${\mathbb{A}=\Z_4}$. 
Let $e$ and $m$ denote the elementary electric and magnetic anyons of $\Z_4$ gauge theory. They obey ${e^4=m^4=1}$ and ${B(e,m)=\ii}$.
$\Z_4$ gauge theory has three gapped boundaries whose corresponding pairs ${(H,[\ga])}$ are ${(1,[1])}$, ${(\Z_2,[1])}$, and ${(\Z_4,[1])}$.
The electric symmetry boundary ${\cL_e=1\oplus e\oplus e^2\oplus e^3}$ is ${(1,[1])}$. Gauging the ${\Z_2<\Z_4}$ subgroup replaces the $\cL_e$ boundary with ${(\Z_2,[1])}$ whose endable anyons are ${1\oplus e^2\oplus m^2\oplus e^2m^2}$. The symmetry described by this new boundary is an anomalous ${\Z_2\times \Z_2}$ symmetry whose anomaly manifests within the SymTFT by the mutual braiding of $e$ and $m$~\cite{CW220303596}. The SymTFT for this anomalous symmetry has been used to show that the deconfined quantum critical point (DQCP) for a ${\Z_2\times \Z_2}$ symmetry is gauge equivalent to an ordinary Landau transition for $\Z_4$ symmetry~\cite{CW220506244, ZL220601222}.
More generally, however, it has also been shown using the SymTFT framework that not all DQCPs of finite symmetries are gauge equivalent to ordinary Landau transitions~\cite{WGS260531601}. 
We note that since a $\Z_4$ and an anomalous ${\Z_2\times\Z_2}$ symmetry are related by gauging, their corresponding SymTFTs are the same after an anyon relabeling. The SymTFT for $\mathbb{Z}_4$ symmetry is $\mathbb{Z}_4$ gauge theory, which maps to twisted ${\Z_2\times\Z_2}$ gauge theory by relabeling 
${e \mapsto m_2}$,
${e^2 \mapsto e_1}$,
${m  \mapsto m_1}$,
and
${m^2 \mapsto e_2}$.

\subsection{Gapless Theories}\label{sect2gapless}

In Sec.~\ref{GroupSectionGappedPhases}, we discussed how gapped phases characterized by the symmetry $\cS$ are described by the gapped physical boundaries of the SymTFT $\fZ(\cS)$. Gapless theories with $\cS$ symmetry are similarly described by gapless physical boundaries of $\fZ(\cS)$. 
The SymTFT provides a useful way of organizing gapless theories according to their symmetry structure.
This is done using the \textit{SymTFT club sandwich} in which a gapped topological interface $\cI$ is inserted between the original SymTFT ${(\Bsym_\cS\mid\fZ(\cS)}$ and a reduced SymTFT ${\fZ(\cS')\mid \Bphys_\cT)}$ with $\Bphys_\cT$ a gapless boundary~\cite{CW220506244, Bhardwaj:2023bbf, BPS240300905}. 
The interface $\cI$ defines a map between the anyons of $\fZ(\cS)$ and those of $\fZ(\cS')$, determining how anyons of $\fZ(\cS)$ are transmitted, identified, split, or confined upon crossing the interface. It is mathematically given by a condensable subgroup, which is a Lagrangian subgroup without the maximality condition ${|L|=|\mathbb{A}|}$. The symmetry $\cS'$ realizes the faithful action of $\cS$ on the gapless boundary $\Bphys_\cT$. Hence, the interface provides a useful separation of the full symmetry data $\cS$ from the symmetry $\cS'$ that characterizes the gapless dynamics.

\begin{marginnote}[]
\entry{SymTFT Club Sandwich}{
\begin{equation*}
\begin{tikzpicture}
\draw [SymTFTcolor,  fill=SymTFTcolor] 
(0,0) -- (0,2) -- (1,2) -- (1,0) -- (0,0) ; 
\draw [SymTFTcolor!55, fill=SymTFTcolor!55] 
(1,0) -- (1,2) -- (2,2) -- (2,0) -- (1,0) ; 
\draw [very thick] (0,0) -- (0,2) ;
\draw [very thick] (2,0) -- (2,2) ;
\draw [very thick] (1,0) -- (1,2) ;
\node at  (1/2,1)  {$\fZ(\cS)$} ;
\node at  (3/2,1)  {$\fZ(\cS')$} ;
\node[above] at (0,2) {$\Bsym_{\cS}$}; 
\node[above] at (2,2) {$\Bphys_{\cT}$}; 
\node[above] at (1,2) {$\cI$}; 
\end{tikzpicture}
\end{equation*}
}
\end{marginnote}

For example, let ${\cS = \Z_4}$ without anomalies. As discussed in Sec.~\ref{sec:Abs},
the SymTFT $\fZ(\Z_4)$ is $\Z_4$ gauge theory and the symmetry boundary is the electric boundary.
Consider the interface $\cI_{e^2m^2}$ on which the condensable subgroup $\{1,e^2m^2\}$ can end topologically.
The reduced SymTFT $\fZ(\cS')$ is twisted $\Z_2$ gauge theory---double-semion topological order---which makes the $\cS'$ symmetry an anomalous $\Z_2$ symmetry.
The interface $\cI_{e^2m^2}$ maps $em$ to $s$ and $em^3$ to $\bar{s}$, the semion and anti-semion, respectively.
Anyons that braid nontrivially with $e^2m^2$, such as $e$ and $m$, cannot cross the interface. This SymTFT club sandwich describes an intrinsically gapless SPT where the $\Z_4$ symmetry acts on the gapless degrees of freedom as an anomalous $\Z_2$ symmetry~\cite{Thorngren:2020wet}. The string order parameter for this SPT is the $e^2m^2$ anyon of $\fZ(\Z_4)$. This anyon ends on the interface $\cI_{e^2m^2}$, while at the electric symmetry boundary it becomes the $m_\mathrm{bdry}^2$ $\Z_4$ symmetry defect through a junction decorated by an $e^2$ charge~\cite{WP220809001, HC231016878, WP231100050, Bhardwaj:2023bbf}.

\section{SymTFT for Categorical Symmetries}\label{Sec:FusionCatSymTFT}

Sec.~\ref{Pedagogicalintro} introduced the SymTFT framework through examples of finite abelian group symmetries. We now extend this discussion to categorical symmetries. We focus on ${(1+1)}$d systems whose symmetry is described by a unitary fusion category $\cS$.
This includes both invertible and non-invertible symmetries. Finite-group symmetries correspond to ${\cS=\Vec_G^\om}$, the category of $G$-graded vector spaces with associator specified by a 3-cocycle ${\om\in \cZ^3(G,\Uone)}$. Here, $G$ may be any finite group, and the cohomology class $[\om]$ encodes its 't~Hooft anomaly. Restricting to abelian $G$ recovers the group symmetries discussed in Sec.~\ref{Pedagogicalintro}. We also discuss extensions to fusion 2-category symmetries. 
Throughout this section, we assume basic familiarity with (braided) fusion categories and refer the reader to~\cite{EGNO} for an introduction.

\subsection{SymTFT for Fusion Category Symmetries}\label{FusionCatSymTFT}

\begin{marginnote}[]
\entry{Drinfeld center}{}
An object of $\cZ(\cS)$ is a pair ${(a,\ga_{-,a})}$, where ${a\in \cS}$ and ${\ga_{x,a}\colon x\otimes a \xrightarrow{\sim} a\otimes x}$ is a half-braiding.
A simple object ${(a,\ga_{-,a})}$ need not have $a$ be simple, and there can be multiple inequivalent $\ga_{-,a}$ for fixed $a$.
Together with $\cS$, the half-braidings determine the fusion rules and $F$ and $R$ symbols of $\cZ(\cS)$~\cite{MT240613438}.
Not every UMTC is a Drinfeld center~\cite{EGNO}.
\end{marginnote}

The SymTFT associated with a unitary fusion category symmetry $\cS$ is a ${(2+1)}$d TFT determined by $\cS$ that admits a gapped boundary that realizes the symmetry $\cS$. 
The Turaev--Viro construction provides precisely such a theory: it uses a fusion category $\cS$ to define a state sum TFT that has a canonical gapped boundary $\Bsym_\cS$ realizing $\cS$~\cite{Turaev:1992hq, Barrett:1993ab, KK11045047}. Denoting this theory by $\TV(\cS)$, we take
\begin{equation}
    \fZ(\cS) = \TV(\cS).
\end{equation}
On a closed spatial surface, its Hilbert space is realized as the ground state space of the Levin--Wen string-net Hamiltonian constructed from $\cS$~\cite{LW0404617, Kirillov:2011mk, Lin:2020bak}. For a finite-group symmetry ${\cS=\Vec_G^\om}$, this construction gives $G$ gauge theory with Dijkgraaf--Witten twist $\om$~\cite{Dijkgraaf:1989pz}, and the canonical symmetry boundary is the Dirichlet boundary. 
Its Hilbert space on a closed spatial surface can be realized as the ground state space of a twisted quantum double model~\cite{Kitaev:1997wr,Hu:2012wx}.
Thus, the Turaev--Viro construction extends the finite abelian group examples of Sec.~\ref{Pedagogicalintro} to general fusion category symmetries.

For the applications considered here, we can work directly with the anyons of $\TV(\cS)$ without requiring an explicit realization of the TFT as proposed in~\cite{Bhardwaj:2023ayw}. This anyon data is encoded by the Drinfeld center $\cZ(\cS)$ of $\cS$, a unitary modular tensor category (UMTC) whose simple objects label the anyons of $\TV(\cS)$~\cite{TV10063501}. 
If $\cS$ admits a nondegenerate braiding, the Drinfeld center ${\cZ(\cS) \simeq \cS\boxtimes\overline{\cS}}$ where $\overline{\cS}$ has the same fusion category as $\cS$ but the reversed braiding.  

\begin{marginnote}[]
\entry{Bulk-to-boundary map}{
$$
\begin{tikzpicture}
\draw [SymTFTcolor,  fill=SymTFTcolor] 
(0,0) -- (0,2) -- (1.75,2) -- (1.75,0) -- (0,0) ; 
\draw [very thick] (0,0) -- (0,2) ;
\draw[thick, red] (0,0.75) -- (1.75,0.75);
\draw[very thick, red] (0,0.75) -- (0,0);
\node[above, red] at (1.75/2,0.75) {$z$};
\node[left, red] at (0,0.75/2) {$F_\cS(z)$};
\node[above, black] at (0,2) {$\Bsym_\cS$}; 
\fill[red] (0, 0.75) circle (0.05cm);
\node at  (1.75/2,1.5)  {$\fZ(\cS)$};
\end{tikzpicture}
$$
}
\end{marginnote}

Bringing the anyons of $\fZ(\cS)$ to the canonical symmetry boundary $\Bsym_\cS$ defines the bulk-to-boundary tensor functor 
\begin{equation}\label{generalBulktoBdyMap}
    F_\cS\colon \ \cZ(\cS)\to \cS,
    \qquad
    (a,\ga_{-,a})\mapsto a.
\end{equation}
The anyon ${z\in\cZ(\cS)}$ gets mapped to ${F_\cS(z) \simeq \bigoplus_{x\in\mathrm{Irr}(\cS)}n_z^x\,x}$, where $n_z^x$ counts the number of independent junction channels through which the anyon $z$ can become the symmetry defect line $x$ on $\Bsym_\cS$.
An anyon ${z}$ can end on $\Bsym_\cS$ only when it can become the trivial defect line $1$ on $\Bsym_\cS$. Thus, $z$ can end on $\Bsym_\cS$ only when ${n_z^1\neq0}$, and $n_z^1$ counts the number of independent ways it can do so.

\noindent{\bf Example: anomaly-free finite $G$ symmetry.} For a general finite group $G$, the simple objects of ${\cS = \Vec_G}$ are ${g\in G}$ with fusion ${g\otimes h \simeq gh}$ and trivial $F$-symbols.
The simple objects of ${\cZ(\Vec_G)}$ are labeled by ${([g], R)}$ with $[g]$ a conjugacy class in $G$ and $R$ an irreducible representation of the centralizer $C_G(g)$. The bulk-to-boundary functor satisfies ${F_{\Vec_G}(([g],R)) \simeq \bigoplus_{h\in[g]}\dim(R)\, h}$~\cite{CCW170704564}. Note how a simple anyon can become a direct sum of symmetry defects at the boundary.
Only the electric anyons ${([1],R)}$ can end on $\Bsym_{\Vec_G}$, so $\Bsym_{\Vec_G}$ is the Dirichlet boundary condition.

\noindent{\bf Example: Ising fusion category symmetry.}
The Ising fusion category ${\cS=\Ising}$ has simple objects ${1,\psi,\sigma}$ with fusion rules ${\psi\otimes\psi\simeq1}$, ${\psi\otimes\sigma\simeq\sigma\otimes\psi\simeq\sigma}$, and ${\sigma\otimes\sigma\simeq1\oplus\psi}$.
This is a non-invertible symmetry.
The Ising fusion category admits a nondegenerate braiding, and its Drinfeld center ${\cZ(\Ising)\simeq\Ising\boxtimes\overline{\Ising}}$~\cite{DGN09060620}. The simple objects of $\cZ(\Ising)$ are labeled by ${a\bar b}$ with ${a,b\in\{1,\psi,\sigma\}}$. Their fusion is componentwise, and their topological spins are ${\theta_{a\bar b}=\theta_a\theta_b^{-1}}$, where $\theta_a$ denotes the topological spin in the chosen braided Ising category.
The bulk-to-boundary functor satisfies ${F_\Ising(a\bar b)\simeq a\otimes b}$. For example, ${F_\Ising(\sigma\bar\sigma)\simeq 1\oplus\psi}$. The anyons that can end on $\Bsym_\Ising$ are $1$, $\psi\bar\psi$, and $\sigma\bar\sigma$.

\begin{marginnote}[]
\entry{Ising fusion categories}{}
The Ising fusion rules admit two inequivalent sets of $F$-symbols, defining two distinct fusion categories denoted $\Ising_\pm$. They differ by the overall sign ${\pm 1}$ of $F_\si^{\si\si\si}$.
These Ising fusion categories are examples of more general Tambara--Yamagami fusion categories.
\end{marginnote}

As discussed in Sec.~\ref{Pedagogicalintro}, different symmetries can have the same SymTFT. That is, there can be two fusion categories $\cS_1$ and $\cS_2$ such that
\begin{equation}
    \TV(\cS_1) = \TV(\cS_2)\qquad \cZ(\cS_1)\simeq\cZ(\cS_2).
\end{equation}
Mathematically, such fusion categories $\cS_1$ and $\cS_2$ are said to be \textit{Morita equivalent}.
This implies that in addition to the canonical gapped boundary condition $\Bsym_{\cS_1}$ realizing an $\cS_1$ symmetry, $\TV(\cS_1)$ also has a gapped boundary condition $\Bsym_{\cS_2}$ realizing an $\cS_2$ symmetry.
Changing between these symmetry boundaries implements a gauging from $\cS_1$ to $\cS_2$.
Hence, the fusion categories of two symmetries $\cS_1$ and $\cS_2$ are Morita equivalent if they are related by a gauging.
A simple example discussed in Sec.~\ref{sec:Abs} is that $\Vec_{\Z_4}$---the fusion category describing an anomaly-free $\Z_4$ symmetry---is Morita equivalent to $\Vec_{\Z_2\times\Z_2}^\om$---the fusion category describing ${\Z_2\times\Z_2}$ symmetry with mixed anomaly $[\om]$. For trivial $[\omega]$, $\Vec_G^\om$ is also Morita equivalent to $\Rep(G)$. These give the Dirichlet and Neumann boundary conditions of the SymTFT, which is a $G$ gauge theory.

\begin{marginnote}[]
\entry{$\Rep(G)$}{}
For a finite group $G$, $\Rep(G)$ is the fusion category of finite-dimensional complex representations of $G$, with intertwining maps as morphisms.
Fusion is the tensor product of representations: ${R_a\!\otimes\! R_b \!=\! \bigoplus_c \!N_{ab}^c R_c}$.
The quantum dimension of $R$ is $\dim(R)$, so $R$ is invertible only when it is one dimensional.
\end{marginnote}

\begin{marginnote}[]
\entry{Group-theoretical}{}
A fusion category $\cS$ is group-theoretical if ${\cZ(\cS)\simeq \cZ(\Vec_G^\om)}$ for some finite group $G$ and 3-cocycle $\om$. Therefore, the SymTFT of a group-theoretical $\cS$ symmetry is a possibly twisted finite group gauge theory. For example, $\Rep(G)$ is group-theoretical while ${\Ising}$ is not.
\end{marginnote}

\subsection{Topological Boundaries and Lagrangian Algebras}
\label{sec:GappedBCLags}

Topological boundaries of $\TV(\cS)$ are classified by \textit{Lagrangian algebras} in the Drinfeld center $\cZ(\cS)$~\cite{Davydov:2010kfz,K13078244}. These objects give the additional structures necessary to specify gapped boundaries for non-abelian UMTCs; in the abelian case we can simply use the Lagrangian subgroups described in Sec.~\ref{sec:Abs}.
As an object of $\cZ(\cS)$, a Lagrangian algebra takes the form
\begin{equation}\label{generalLagAlgObj}
    \cL=\bigoplus_{z\in\mathrm{Irr}(\cZ(\cS))}n_z\,z,
\end{equation}
and we denote its corresponding topological boundary by $\fB_\cL$. The corresponding Lagrangian algebra for the canonical symmetry boundary $\Bsym_\cS$ is ${\cL_\cS=\bigoplus_z n_z^1\,z}$, with multiplicities determined by the bulk-to-boundary functor~\eqref{generalBulktoBdyMap}. 
The algebra structure on $\cL$ encodes the fusion of endpoints of lines ending on $\fB_\cL$. It consists of a multiplication morphism ${M\colon\cL\otimes\cL\to\cL}$ compatible with the associator and braiding of $\cZ(\cS)$.
To be a Lagrangian algebra, $\cL$ must satisfy other conditions, including ${n_1 = 1}$ and ${d_\cL \equiv \sum_z n_z d_z=\cD}$, where $d_z$ is the quantum dimension of $z$ and ${\cD = \sqrt{\sum_z d_z^2}}$ is the total quantum dimension. (See~\cite{Davydov:2010kfz, CCW170704564} for a complete definition.)

Physically, $\cL$ describes the bulk anyons that can end on $\fB_\cL$, and $n_z$ counts the number of inequivalent ways the anyon $z$ can end.
The requirement ${n_1 = 1}$ means that the only intrinsic local topological operator on $\fB_\cL$ is the identity.
Maximality ${d_\cL=\cD}$ ensures that condensation leaves no nontrivial deconfined bulk anyons, so $\fB_\cL$ describes a boundary to the vacuum.
Compatibility with bulk braiding requires each anyon $z$ with ${n_z\neq 0}$ in $\cL$ to be a boson. In simple cases, the object $\cL$ uniquely determines the boundary $\fB_\cL$. However, in general, there can be two Lagrangian algebras with the same underlying object but inequivalent algebra structures~\cite{D09081044, KB250103314, GSW260531602}.
The topological defect lines on $\fB_\cL$ form a fusion category $\cS_\cL$. Hence, $\fB_\cL$ realizes an $\cS_\cL$ symmetry. For the Lagrangian algebra $\cL_\cS$, this is the $\cS$ symmetry. 
\begin{marginnote}[]
\entry{Boundary fusion category}{}
The fusion category $\cS_\cL$ is described by the right $\cL$-modules in $\cZ(\cS)$~\cite{KK11045047,K13078244}. 
Its fusion uses the relative tensor product
$\otimes_\cL$, whose tensor unit is $\cL$~\cite{K13078244}.
The corresponding bulk-to-boundary functor $F_\cL$ satisfies ${F_\cL(z)=z\otimes\cL}$.
\end{marginnote}

\noindent{\bf Example: finite $G$ and $\Rep(G)$ symmetries.}
Consider ${\cS = \Vec_G}$ and the canonical topological boundary $\Bsym_{\Vec_G}$ of its SymTFT $\fZ(\Vec_G)$.
From Sec.~\ref{FusionCatSymTFT}, $\Bsym_{\Vec_G}$ is the Dirichlet boundary condition with bulk-to-boundary functor ${F_{\Vec_G}(([g],R)) \simeq \bigoplus_{h\in[g]}\dim(R)\, h}$. Thus, its corresponding Lagrangian algebra is ${\cL_{\Vec_{G}}
    =
    \bigoplus_{R}
        \,\dim(R)\,([1],R)}$.
The SymTFT $\fZ(\Vec_G)$ always has at least one other topological boundary, the Lagrangian algebra of which is ${\cL_{\Rep(G)}=
    \bigoplus_{[g]}([g],1)}$.
Its symmetry boundary $\fB_{\cL_{\Rep(G)}}$ realizes a $\Rep(G)$ symmetry, which is non-invertible when $G$ is non-abelian. 
The corresponding bulk-to-boundary map satisfies ${F_{\Rep(G)}(([g],R)) = \mathrm{Ind}^G_{C_G(g)}(R)}$~\cite{D09081044, CCW170704564}.
Whereas electric charges ${([1],R)}$ can end on $\fB_{\cL_{\Vec_G}}$, magnetic flux ${([g],1)}$ can end on $\fB_{\cL_{\Rep(G)}}$. 
Changing the symmetry boundary from $\fB_{\cL_{\Vec_G}}$ to $\fB_{\cL_{\Rep(G)}}$ implements gauging the  entire $G$ symmetry.

More generally, Lagrangian algebras in $\cZ(\Vec_G)$ are classified by pairs ${(H,[\ga])}$, where ${H\leq G}$ and ${[\ga]\in \cH^2(H,\Uone)}$, modulo conjugation by $G$, ${(H,[\gamma])\sim (gHg^{-1},[\gamma^g])}$~\cite{D09081044, beigi2011, GSW260531602}.
Starting from the $\fB_{\cL_{\Vec_G}}$ boundary, the boundary labeled by ${(H,[\ga])}$ is obtained by stacking with the $H$-SPT phase $[\ga]$ and then gauging $H$.

\noindent{\bf Example: Ising fusion category symmetry.}
For the SymTFT $\fZ(\Ising)$ discussed in
Sec.~\ref{FusionCatSymTFT}, the bulk-to-boundary functor satisfies ${F_\Ising(a\bar b)\simeq a\otimes b}$. Thus, its corresponding Lagrangian algebra is ${\cL_\Ising
    =1\oplus\psi\bar\psi\oplus\sigma\bar\sigma}$.
Up to isomorphisms, this is the only Lagrangian algebra in $\cZ(\Ising)$.

\subsection{Generalized Charges}\label{GenChargeSubSecFusionCats}

For an anomaly-free finite $G$ symmetry, operators at the end of a ${g\in G}$ symmetry defect transform in representations of its centralizer $C_G(g)$.
Genuine local operators transform in representations of ${C_G(1) = G}$.
The ${h\in G}$ symmetry transformation sends $g$ to $hgh^{-1}$, so a complete symmetry multiplet includes all defect sectors in the conjugacy class $[g]$.
Thus, operators in a theory with $G$ symmetry organize into multiplets labeled by pairs ${([g],R)}$ called \textit{generalized charges}, where $R$ is an irreducible representation of $C_G(g)$.
The generalized charges of a general fusion category symmetry $\cS$ are labeled by simple objects of $\cZ(\cS)$, or equivalently anyons of $\fZ(\cS)$~\cite{LOS220805495, Bhardwaj:2023ayw}.

The correlation functions of generalized charges in a theory $\cT$ with $\cS$ symmetry are encoded in the physical boundary $\Bphys_\cT$ of the SymTFT sandwich ${(\Bsym_{\cS}\mid\fZ(\cS)\mid\Bphys_\cT)}$.
Consider the bulk anyon $z$ of the SymTFT sandwich which becomes ${F_\cS(z)\simeq \bigoplus_{x} n_z^x\,x}$ on the symmetry boundary $\Bsym_{\cS}$. The generalized charge $z$ is realized in $\cT$ if and only if the anyon $z$ can end on the physical boundary $\Bphys_\cT$. (The physical boundary endpoint need not be topological.)
Suppose the anyon $z$ can end on $\Bphys_\cT$, and extends across $\fZ(\cS)$ to $\Bsym_{\cS}$.
This corresponds to a local operator of $\cT$ attached to an ${F_\cS(z)\simeq \bigoplus_{x} n_z^x\,x}$ symmetry defect line.
In the ${x=1}$ component of $F_\cS(z)$, this is a genuine local operator; otherwise it is a non-genuine operator (sometimes known as a string order parameter or disorder operator).
For non-invertible symmetries, a single irreducible multiplet can contain both genuine and non-genuine local operators.

Consider a local operator $\cO_z$ with generalized charge $z$. 
In the SymTFT sandwich, a closed loop of a bulk simple anyon $z'$ can act on this multiplet by linking with $z$. This defines an action of $z'$ on $\cO_z$ by
\begin{equation}
    z'\cdot\cO_z = \frac{S_{z'z}}{S_{1z}}\,\cO_z,
\end{equation}
where $S$ is the modular $S$ matrix of $\cZ(\cS)$~\cite{K0506438}. In the abelian case, the charge given by the coefficient on the right-hand side above is a phase. 
For a genuine local operator $\cO_z$, moving the $z'$ anyon loop to $\Bsym_\cS$ identifies ${z'\cdot\cO_z}$ with the action of $F_\cS(z')$ on $\cO_z$.
The defect $F_\cS(z')$ is generally a direct sum of simple symmetry defects, and this equation gives the action of the entire sum.
When $\cO_z$ is non-genuine, ${z'\cdot \cO_z}$ does not have this interpretation since moving the $z'$ anyon loop to $\Bsym_\cS$ requires it to cross the attached symmetry defect, and its boundary realization then depends on both $F_\cS(z')$ and this crossing data. 
More general networks of bulk anyons, coupled through junctions to symmetry-boundary defects, can act on $z$ and change the twist sector of $\cO_z$ while
preserving its generalized charge~\cite{CLS180204445,LOS220805495}.

\begin{marginnote}[]
\entry{Generalized charge  multiplet}{
$$
\begin{tikzpicture}
\begin{scope}[shift={(0,0)}]
\draw [SymTFTcolor,  fill=SymTFTcolor] 
(0,0) -- (0,2) -- (1.25,2) -- (1.25,0) -- (0,0) ; 
\draw [white] (0,0) -- (0,2) -- (1.25,2) -- (1.25,0) -- (0,0)  ; 
\draw [very thick] (0,0) -- (0,2) ;
\draw [very thick] (1.25,0) -- (1.25,2) ;
\draw [red, thick] (0,1) -- (1.25,1);
\draw [red, very thick] (0,1) -- (0,0);
\node at  (1.25/2,0.5)  {$\fZ(\cS)$} ;
\node[red, above] at (1.25/2,1) {$z$};
\node[above] at (0.125,2) {$\Bsym_\cS$}; 
\node[above] at (1.25,2) {$\Bphys_\cT$}; 
\fill[red] (0,1) circle (0.05cm);
\fill[red] (1.25,1) circle (0.05cm);
\end{scope}
\begin{scope}[shift={(2,0)}]
\node at (-0.4,1) {$=$};
\draw [very thick] (0,0) -- (0,2) ;
\fill[red] (0,1) circle (0.05cm);
\node[right] at (0,1) {$\cO_{z}$};
\draw [red, very thick] (0,1) -- (0,0);
\end{scope}
\end{tikzpicture}
$$
}
\end{marginnote}

\noindent{\bf Example: $\Vec_G$ and $\Rep(G)$ symmetries.}
These symmetries have the same generalized charges $([g],R)$ (since they are Morita equivalent), but their respective symmetry boundaries organize them differently. 
For a $\Vec_G$ symmetry, genuine operators have charges ${([1],R)}$ and transform under ${g\in G}$ by the matrix $R(g)$. A bulk flux loop ${([h],1)}$ when moved to the symmetry boundary becomes the object ${\bigoplus_{k\in[h]}k}$ of $\Vec_G$. It acts on the ${([1],R)}$ multiplet as ${[h]\cdot\cO_{R} = \frac{|[h]|\chi_R(h)}{\dim R}\, \cO_{R}}$, where $\chi_R$ is the character of $R$.
For $\Rep(G)$ symmetry, a bulk charge loop ${([1],R)}$ becomes the simple object $R$ of $\Rep(G)$ on the symmetry boundary.
It acts on the ${([g],1)}$ multiplet by ${R\cdot\cO_{[g]}=\chi_R(g)\,\cO_{[g]}}$.

\noindent{\bf Example: Ising fusion category symmetry.}
Among the nine Ising generalized charges, $1$, $\psi\bar\psi$, and $\sigma\bar\sigma$ are the only ones including genuine local operators.
Since $a\bar{b}$ becomes ${a\otimes b}$ on the symmetry boundary, the $\Ising$ symmetry defect $a$ can be represented by the bulk anyon ${a\bar1}$. Using the $S$ matrix of $\fZ(\Ising)$, it acts on the genuine local operators by the scalars
\begin{equation}\label{IsingChargeAction}
    \begin{array}{c|ccc}
        & ~\cO_1~ & ~\cO_{\psi\bar\psi}~
        & ~\cO_{\sigma\bar\sigma}~ \\ \hline
        \psi   & 1 & 1 & -1 \\
        \sigma & \sqrt2 & -\sqrt2 & 0
    \end{array}.
\end{equation}
Since ${F_\Ising(\sigma\bar\sigma)\simeq1\oplus\psi}$, the $\sigma\bar\sigma$ multiplet contains both a genuine and $\psi$-twisted local operator. 
With an appropriate junction, a $\si$ symmetry defect can thus map a genuine local operator in the $\sigma\bar\sigma$ multiplet to a $\psi$-twisted operator in the same multiplet. In the Ising CFT, this action implements Kramers--Wannier duality, exchanging the order and disorder operators~\cite{KC1971, FFR0404051}.

\subsection{Classification of Gapped Phases}
\label{fusionCatGappedPhaseSec}

Gapped ${(1+1)}$d phases with $\cS$ symmetry, including SSB and SPT phases, are described in the deep IR by $\cS$-symmetric ${(1+1)}$d TFTs. Such TFTs are realized by SymTFT sandwiches ${(\Bsym_\cS\mid\fZ(\cS)\mid\Bphys)}$ with a topological physical boundary $\Bphys$. Fixing the symmetry boundary $\Bsym_\cS$, gapped phases with $\cS$ symmetry are therefore classified by Lagrangian algebras $\mathcal{L}$ in $\cZ(\cS)$~\cite{KLW200514178,Garre-Rubio:2022uum,Bhardwaj:2023ayw, BBP231003784}. {This is equivalent to classifying module categories over $\cS$~\cite{TW191202817, Inamura:2021szw}.}

Whether a gapped phase is an SSB or SPT phase depends on which generalized charges---order parameters---are realized in its IR TFT. 
Suppose ${\Bphys = \fB_\cL}$ with ${\cL \simeq \bigoplus_z n_z^\mathrm{phys} z}$, and write ${\cT_\cL = (\Bsym_\cS\mid\fZ(\cS)\mid\fB_\cL)}$.
An anyon $z$ can end on $n_z^\mathrm{phys}$ independent topological operators of $\Bphys$.
From the $\Bsym_\cS$ bulk-to-boundary functor ${F_\cS(z)\simeq\bigoplus_x n_z^x\,x}$, this implies that the number of independent topological operators at the end of an $x$ symmetry defect in ${\cT_\cL}$ is ${N_x=\sum_z n_z^x\, n_z^\mathrm{phys}}$.
The ${x=1}$ channels give genuine local operators---local order parameters---while the ${x\neq1}$ channels give non-genuine local operators---string order parameters. 
The local order parameters distinguish different ground states. Consequently, the number of ground states of $\cT_\cL$ is~\cite{LWW14086514}
\begin{equation}\label{GappedVacuumCount}
    N_{\mathrm{gs}}=N_1
    =\sum_z n_z^1\, n_z^\mathrm{phys}
    =\dim\Hom(\cL_\cS,\cL).
\end{equation}
If ${\Hom(\cL_\cS,\cL)}$ is one-dimensional, then $\cT_\cL$ describes an $\cS$ SPT phase.
Otherwise, ${N_{\mathrm{gs}}>1}$ and $\cT_\cL$ describes an SSB phase.

\noindent{\bf Example: $\Vec_G$ and $\Rep(G)$ symmetries.}
As discussed in Sec.~\ref{sec:GappedBCLags}, topological boundaries of untwisted $G$ gauge theory are labeled by pairs ${(H,[\ga])}$, where ${H\leq G}$ and ${[\ga]\in\cH^2(H,\Uone)}$, modulo $G$ conjugation. Denote the corresponding Lagrangian algebra by $\cL_{H,\ga}$. 
With symmetry boundary $\Bsym_{\Vec_G}$ and physical boundary $\fB_{\cL_{H,\ga}}$, the SymTFT sandwich describes a phase where $G$ spontaneously breaks to $H$ and the unbroken $H$ symmetry has the SPT class $[\ga]$. Its ground states are labeled by ${G/H}$, which implies ${N_\mathrm{gs}^{\Vec_G}(H,[\ga]) = |G/H|}$. 
Replacing the symmetry boundary by $\Bsym_{\Rep(G)}$ while keeping the physical boundary fixed implements gauging $G$. The resulting $\Rep(G)$-symmetric phase labeled by ${(H,[\ga])}$ is described by ${(1+1)}$d $G$ gauge theory Higgsed to $H$ with Dijkgraaf--Witten twist $[\ga]$ for the residual $H$ gauge theory~\cite{Inamura:2021szw}.
Its ground states are therefore labeled by irreducible $\ga$-projective representations of $H$, which implies ${N_\mathrm{gs}^{\Rep(G)}(H,[\ga]) = |\mathrm{Irr}_\ga(H)|}$~\cite{T07060160}.

\noindent{\bf Example: Ising fusion category symmetry.}
As discussed in Sec.~\ref{sec:GappedBCLags}, $\cZ(\Ising)$ has only one Lagrangian algebra ${\cL_\Ising=1\oplus\psi\bar\psi\oplus\sigma\bar\sigma}$.
With the symmetry boundary fixed to $\Bsym_\Ising$, the only topological physical boundary is also described by $\cL_\Ising$.
This implies that the $\Ising$ symmetry has only one gapped phase, which is an SSB phase that, from~\eqref{GappedVacuumCount}, has ${N_\mathrm{gs}^\Ising=3}$ ground states. Let $\cP_1$, $\cP_\psi$, and $\cP_\si$ denote the projectors onto these three ground states.
It follows from~\eqref{IsingChargeAction} that the $\Ising$ symmetry acts on them as~\cite{BBP231003784}
\begin{equation}
    \begin{gathered}
        \psi\cdot \cP_1 = \cP_\psi,\qquad 
        \psi\cdot \cP_\psi = \cP_1,\qquad 
        \psi\cdot \cP_\si = \cP_\si,\\
        \si\cdot \cP_1 = \frac1{\sqrt{2}}\cP_\si,\qquad 
        \si\cdot \cP_\psi = \frac1{\sqrt{2}}\cP_\si,\qquad 
        \si\cdot \cP_\si = \sqrt{2}(\cP_1+\cP_\psi).
    \end{gathered}
\end{equation}

In general, the physical boundary $\Bphys$ can always be taken as the symmetry boundary $\Bsym_\cS$. The corresponding SymTFT sandwich realizes a ${(1+1)}$d TFT that maximally spontaneously breaks the $\cS$ symmetry and has ${\dim\Hom(\cL_\cS,\cL_\cS) = |\mathrm{Irr}(\cS)|}$ ground states, one for each simple object of $\cS$.

\subsection{Gapless Boundaries and Condensable Algebras}

A gapless theory with $\cS$ symmetry is described by a SymTFT sandwich whose physical boundary $\Bphys$ is gapless. As discussed in Sec.~\ref{sect2gapless}, its symmetry structure can be organized using a SymTFT club sandwich, with a topological interface separating $\fZ(\cS)$ from a reduced SymTFT $\fZ(\cS')$~\cite{CW220506244,Bhardwaj:2023bbf}.
Here, we consider a topological condensation interface $\cI_\cA$ specified by a \textit{condensable algebra} ${\cA\in \cZ(\cS)}$, which satisfies the same algebraic requirements as a Lagrangian algebra in Sec.~\ref{sec:GappedBCLags}, except that maximality is not required~\cite{K13078244}. As an object, it takes the form
\begin{equation}
    \cA = \bigoplus_{z\in\mathrm{Irr}(\cZ(\cS))}n^\cA_z\,z.
\end{equation}
Physically, a condensable algebra $\cA$ specifies which anyons of $\cZ(\cS)$ can end on the topological interface $\cI_\cA$, and $n_z^{\cA}$ gives the number of independent ways $z$ can end on $\cI_\cA$.
When ${d_\cA = \cD}$, $\cA$ is a Lagrangian algebra and the reduced SymTFT $\fZ(\cS')$ is trivial.

Fusing $\cI_\cA$ with $\Bsym_\cS$ gives a topological boundary of $\fZ(\cS')$ with $\cS$ symmetry.
This boundary of $\fZ(\cS')$ can have genuine local topological operators arising from anyon lines that can end topologically on both $\Bsym_\cS$ and $\cI_\cA$. As in Sec.~\ref{fusionCatGappedPhaseSec}, the number of such topological operators is ${n_{\cS,\cA}=\dim\Hom(\cL_\cS,\cA)}$.
For ${n_{\cS,\cA}>1}$, this fused boundary decomposes into a direct sum of simple topological boundaries of $\fZ(\cS')$~\cite{Bhardwaj:2023bbf, BPS240300905}.

Condensable algebras organize gapless theories into symmetry-based classes. Consider a SymTFT club sandwich with symmetry boundary $\Bsym_\cS$ and topological interface $\cI_\cA$ such that only the trivial anyon of $\fZ(\cS')$ ends topologically on the physical boundary. The corresponding gapless theory has $n_{\cS,\cA}$ topological local operators and, therefore, $n_{\cS,\cA}$ gapless superselection sectors.
A gapless theory with ${n_{\cS,\cA}=1}$ is a \textit{gapless SPT} (gSPT)~\cite{SPV170501557, BPS240300905}. 
Each symmetry $\cS$ has a canonical gSPT with ${\cA=1}$ that putatively corresponds to a CFT that realizes every generalized charge in the IR. Other gSPTs have generalized charges absent from the IR, which causes $\cS$ to act as a reduced symmetry $\cS'$.
A gapless theory with ${n_{\cS,\cA}>1}$ is a \textit{gapless SSB} (gSSB). The $n_{\cS,\cA}$ superselection sectors arise from the spontaneous breaking of $\cS$.
A gSPT or gSSB is \textit{intrinsically gapless} (igSPT or igSSB) \cite{BPS240300905} if opening a gap while preserving $\cS$ and keeping the generalized charges already absent from the IR gapped necessarily increases the number of superselection sectors. Thus, gapping an igSPT requires SSB, whereas gapping an igSSB requires further SSB. These obstructions are naturally encoded in the partial order, and associated Hasse diagram, of condensable algebras~\cite{BPS240300905}.

\noindent{\bf Example: $\Vec_G$ and $\Rep(G)$ symmetries.}
The condensation interfaces $\cI_\cA$ of untwisted $G$ gauge theory are classified by ${(H,N,\ga,\eps)}$ where ${H \leq G}$, ${N\triangleleft H}$, ${\ga\in\cZ^2(N,\Uone)}$, and ${\eps\colon H\times N \to \Uone}$; the map $\eps$ is not arbitrary and must satisfy certain compatibility conditions~\cite{Davydov:2016lag}. The computation of the corresponding condensable algebras is automated in Ref.~\cite{Warman:2026eml}. An example with a nontrivial $N$ and $\epsilon$ is discussed in Sec.~\ref{sect2gapless}, where $\epsilon$ dresses the $N$ fluxes $m^2$ with charge $e^2$, resulting in a $\fZ(\cS')$ with a nontrivial Dijkgraaf--Witten twist. For a general $G$, for simplicity, consider the topological interface
$\cI_{\cA_H}$ that Higgses $G$ to a proper subgroup ${H<G}$. 
The reduced SymTFT $\fZ(\cS')$ is untwisted $H$ gauge theory.
The corresponding condensable algebra is ${\cA_H=\bigoplus_{R\in\mathrm{Irr}(G)}m_R\,([1],R)}$, where $m_R$ is the multiplicity of the trivial representation in the restriction of $R$ to $H$.
Thus, an electric anyon $([1],R)$ of $G$ gauge theory can end topologically on $\cI_{\cA_H}$ when the restriction of $R$ to $H$ contains the trivial representation.
Choosing the symmetry boundary $\Bsym_{\Vec_G}$, the SymTFT club sandwich has ${n_{\Vec_G,\cA_H}=|G/H|>1}$ superselection sectors. The resulting gapless theory is a gSSB with $G$ spontaneously broken to $H$.
Replacing the symmetry boundary by $\Bsym_{\Rep(G)}$ gauges $G$, which maps this gSSB system to a $\Rep(G)$ gSPT system since ${n_{\Rep(G),\cA_H}=1}$.

\noindent{\bf Example: Ising fusion category symmetry.}
The SymTFT $\fZ(\Ising)$ has one nontrivial, non-Lagrangian condensable algebra. It is given by ${\cA=1\oplus\psi\bar\psi}$. The corresponding interface $\cI_\cA$ reduces $\fZ(\Ising)$ to untwisted $\Z_2$ gauge theory $\fZ(\Vec_{\Z_2})$, and fusing $\cI_\cA$ with $\Bsym_\Ising$ gives the direct sum of its electric and magnetic boundaries~\cite{Bhardwaj:2023bbf}.
Since ${\cL_\Ising=1\oplus\psi\bar\psi\oplus\sigma\bar\sigma}$, we have ${n_{\Ising,\cA}=2}$, so the club sandwich describes a gSSB with two superselection sectors.
For example, choosing $\Bphys$ to realize the Ising CFT $\cT_{\mathrm{Ising}}$ gives ${\cT_{\mathrm{Ising}}\oplus(\cT_{\mathrm{Ising}}/\Z_2)\simeq \cT_{\mathrm{Ising}}\oplus \cT_{\mathrm{Ising}}}$. The non-invertible defect $\si$ exchanges the two gapless sectors.
As discussed in Sec.~\ref{fusionCatGappedPhaseSec}, there is only one gapped phase with $\Ising$ symmetry and it has 3 ground states. Thus, opening a gap requires an increase in the number of superselection sectors,
making this an igSSB.

\subsection{Anomalies}

't~Hooft anomalies constrain RG flows by obstructing a symmetry-preserving gapped phase with a unique ground state on every closed spatial manifold.
Consider an $\cS$-preserving RG flow from an ultraviolet (UV) theory with fusion category symmetry $\cS$ to an IR theory with fusion category symmetry $\cS_\IR$.
The RG flow defines a map between the UV and IR symmetry categories described by a tensor functor
\begin{equation}\label{UVtoIRsymMap}
    F_\mathrm{RG}\colon \cS \to \cS_\IR. 
\end{equation}
When ${\cS = \Vec_{G}^{\om}}$ and ${\cS_\IR = \Vec_{G_{\IR}}^{\om_{\IR}}}$, this functor induces a group homomorphism ${\rho\colon G\to G_\IR}$ whose pullback $\rho^*$ satisfies the anomaly-matching condition ${\rho^*[\om_\IR] = [\om]}$~\cite{EGNO}.
A symmetry-preserving gapped phase with a unique ground state has IR symmetry ${\cS_\IR = \Vec}$. Therefore, an $\cS$ symmetry is anomaly-free if $\cS$ admits a \textit{fiber functor}, a tensor functor ${F\colon \cS\to\Vec}$; otherwise, the $\cS$ symmetry is anomalous. Inequivalent fiber functors classify inequivalent $\cS$ SPT phases~\cite{TW191202817, I210315588}.

In the SymTFT, the tensor functor~\eqref{UVtoIRsymMap} is realized by an interface between symmetry boundaries $\Bsym_{\cS}$ and $\Bsym_{\cS_\IR}$. 
This interface extends into the bulk as a topological interface $\cI_F$ separating $\fZ(\cS)$ and $\fZ(\cS_\IR)$ which must satisfy~\cite{Antinucci:2025fjp}
\begin{equation}\label{MatchingEquation}
    \Bsym_{\cS}\times\cI_F
    \simeq\Bsym_{\cS_\IR}.
\end{equation}
If the $\cS$ symmetry is anomaly-free, the fiber functor ${F:\cS\to\Vec}$ gives a topological interface $\cI_F$ between $\fZ(\cS)$ and $\fZ(\Vec)$.
Since $\fZ(\Vec)$ is the trivial bulk theory, $\cI_F$ is a topological boundary of $\fZ(\cS)$. The condition ${\Bsym_\cS\times\cI_F\simeq\Bsym_\Vec}$ requires that only the trivial anyon of $\fZ(\cS)$ can end topologically on both $\Bsym_\cS$ and $\cI_F$. 
Hence, $\cS$ is anomaly-free if and only if there exists
a Lagrangian algebra $\cL$ satisfying ${\dim\Hom(\cL_\cS,\cL)=1}$~\cite{KLW200514178,ZC230401262}.
By Eq.~\eqref{GappedVacuumCount}, this is precisely the condition for the sandwich ${(\Bsym_\cS\mid\fZ(\cS)\mid\fB_\cL)}$ to have a unique ground state and correspond to an $\cS$ SPT. This SymTFT-based approach was used in Refs.~\cite{Cordova:2023bja,Antinucci:2023ezl} to diagnose anomalies of non-invertible symmetries in ${(3+1)}$d.

\begin{marginnote}[]
\entry{SymTFT RG-Quiche}{
$$
\begin{tikzpicture}
\draw [SymTFTcolor!55, fill=SymTFTcolor!55] 
(0,0) -- (0,1) -- (1.75,1) -- (1.75,0) -- (0,0) ; 
\draw [SymTFTcolor,  fill=SymTFTcolor] 
(0,1) -- (0,2) -- (1.75,2) -- (1.75,1) -- (0,1) ; 
\draw [very thick] (0,0) -- (0,2) ;
\draw [very thick] (0,1) -- (1.75,1) ;
\node at  (1.75/2,1.5)  {$\fZ(\cS)$} ;
\node at  (1.75/2,0.5)  {$\fZ(\cS_\IR)$} ;
\node[above] at (0,2) {$\Bsym_{\cS}$}; 
\node[below] at (0,0) {$\Bsym_{\cS_\IR}$}; 
\node[right] at (1.75,1) {$\cI_F$}; 
\filldraw[fill=white, draw=black]
  (0,1) ++(-0.05cm,-0.05cm) rectangle ++(0.1cm,0.1cm);
\end{tikzpicture}
$$
}
\end{marginnote}

The fiber functor criterion is a coarse diagnostic determining whether an $\cS$ symmetry is anomalous.
The anomalies of fusion category symmetries can be more finely characterized in terms of \textit{anomalous simple categories} (ASCies)~\cite{Antinucci:2025fjp}, which capture irreducible building blocks of categorical anomalies.
Consider a short exact sequence of fusion categories connected by tensor functors
\begin{equation}\label{SEE}
    \cN
    \xrightarrow{~I~}
    \cS
    \xrightarrow{~P~}
    \cQ,\qquad \mathrm{im}(I)=\ker(P),
\end{equation}
where $I$ embeds $\cN$ into $\cS$ and $P$ is surjective and normal. $\cN$ is a normal subcategory of $\cS$, which implies that it admits a fiber functor and the $\cN$ subsymmetry of $\cS$ is anomaly-free. 
The tensor functor $P$ describes a UV-to-IR symmetry map where the $\cN$ subsymmetry trivializes in the IR and the UV $\cS$ symmetry becomes a $\cQ$ symmetry.
If the quotient $\cQ$ has no nontrivial normal subcategories, it is called an \textit{anomalous simple category} (ASCy).
A fusion category $\cS$ can have multiple inequivalent ASCies, and their collection provides a refined characterization of its anomaly. 
Some examples of low-rank ASCies are $\Vec$, $\Vec^{\omega}_{\Z_2}$ with nontrivial ${[\om]}$, the Fibonacci fusion category $\mathsf{Fib}$, $\Ising$, and the Haagerup fusion category $\cH_3$. Several of these do contain a subcategory that admits a fiber functor, such as ${\Vec_{\Z_2} \subset \Ising}$ or ${\Vec_{\Z_3}\subset \cH_3}$, but these are not normal subcategories. 
In general, calculating the ASCies of a fusion category is technically difficult, and the SymTFT provides a useful, physical realization of~\eqref{SEE} with which these quotients can be determined.

\begin{marginnote}[]
\entry{SymTFT quiche for the short exact sequence~\eqref{SEE}}{
$$
\begin{tikzpicture}
\draw [SymTFTcolor!55, fill=SymTFTcolor!55] 
(0,-1) -- (0,0) -- (1.3,0) -- (1.3,-1) -- (0,-1) ; 
\draw [SymTFTcolor!100, fill=SymTFTcolor!100] 
(0,0) -- (0,1) -- (1.3,1) -- (1.3,0) -- (0,0) ; 
\draw [SymTFTcolor!150,  fill=SymTFTcolor!150] 
(0,1) -- (0,2) -- (1.3,2) -- (1.3,1) -- (0,1) ; 
\draw [very thick] (0,-1) -- (0,2) ;
\draw [very thick] (0,1) -- (1.3,1) ;
\draw [very thick] (0,0) -- (1.3,0) ;
\node at  (1.3/2,1.5)  {$\fZ(\cN)$} ;
\node at  (1.3/2,0.5)  {$\fZ(\cS)$} ;
\node at  (1.3/2,-0.5)  {$\fZ(\cQ)$} ;
\node at (-0.38,1.5) {$\Bsym_{\cN}$}; 
\node at (-0.38,0.5) {$\Bsym_{\cS}$}; 
\node at (-0.38,-0.5) {$\Bsym_{\cQ}$}; 
\node[right] at (1.3,1) {$\cI_I$}; 
\node[right] at (1.3,0) {$\cI_P$}; 
\filldraw[fill=white, draw=black]
  (0,1) ++(-0.05cm,-0.05cm) rectangle ++(0.1cm,0.1cm);
\filldraw[fill=white, draw=black]
  (0,0) ++(-0.05cm,-0.05cm) rectangle ++(0.1cm,0.1cm);
\end{tikzpicture}
$$
}
\end{marginnote}

\noindent{\bf Example: finite $G$ symmetry.}
For ${\cS=\Vec_G^\om}$, a fiber functor exists only when ${[\om]}$ is trivial, reproducing the usual group-cohomology anomaly criterion~\cite{EGNO,TW191202817}.
The possible ASCies of an anomalous $G$ symmetry are anomalous ${G/N}$ symmetries whose anomaly classes pull back to $[\om]$~\cite{Antinucci:2025fjp}.
For example, consider the anomalous $\Z_8$ symmetry with ${\cS = \Vec_{\Z_8}^{\om}}$ with ${[\om]= 4\in\cH^3(\Z_8,\Uone)\cong\Z_8}$. It fits into two short exact sequences of tensor functors
\begin{equation}
    \Vec_{\Z_2}
    \longrightarrow
    \Vec_{\Z_8}^{\om}
    \xrightarrow{~P_\pm~}
    \Vec_{\Z_4}^{\om_\pm}.
\end{equation}
Here, $\Vec_{\Z_2}$ is a maximal normal subcategory, and the two quotient categories $\Vec_{\Z_4}^{\om_\pm}$ with ${[\om_\pm] = \pm 1\in \cH^3(\Z_4,\Uone)}$ are inequivalent ASCies. Therefore, there are two ways to trivialize the anomaly-free ${\Z_2<\Z_8}$ symmetry and they lead to two different anomalous ${\Z_8/\Z_2\cong \Z_4}$ symmetries in the IR.

\noindent{\bf Example: ${\mathsf{TY}(\Z_4)}$ symmetry.}
Consider the Tambara--Yamagami fusion category ${\cS = \mathsf{TY}(\Z_4)}$ with bicharacter ${\chi(a,b)=i^{ab}}$ and Frobenius–Schur indicator ${\eps=+1}$.
It does not admit a fiber functor and, therefore, is an anomalous fusion category symmetry~\cite{TW191202817, ZC230401262}.
It has an anomaly-free $\Z_2$ subsymmetry, and the corresponding $\Vec_{\Z_2}$ subcategory can occur as $\cN$ in~\eqref{SEE} giving two inequivalent ASCies~\cite{Antinucci:2025fjp}:
\begin{equation}
    \Vec_{\Z_2}
    \xrightarrow{~I~}\mathsf{TY}(\Z_4)
    \xrightarrow{~P_+~}\Vec_{\Z_4}^{\om_+}
    \qquad
    \text{and}
    \qquad
    \Vec_{\Z_2}
    \xrightarrow{~I~}\mathsf{TY}(\Z_4)
    \xrightarrow{~P_-~}\Vec_{\Z_2\times\Z_2}^{\om_-}.
\end{equation}
Here, ${[\om_+]=1\in\cH^3(\Z_4,\Uone)\cong\Z_4}$ and $[\om_-]$ corresponds to the mixed anomaly of ${\Z_2\times\Z_2}$.
Therefore, there are two ways to trivialize the anomaly-free ${\Z_2}$ symmetry of $\mathsf{TY}(\Z_4)$ and they lead to two different anomalous grouplike symmetries in the IR.

\subsection{SymTFT for Fusion 2-Category Symmetries}\label{sec:Gapped2plus1d}

Fusion category symmetries in ${(1+1)}$d generalize to fusion $d$-category symmetries in ${(d+1)}$d.
The objects of a fusion $d$-category $\cS$ represent 0-form symmetry defects, while its $n$-morphisms describe codimension-${(n+1})$ symmetry defects and junctions.
This structure incorporates higher-form symmetries and interplays between defects of different codimensions, such as higher-group symmetries.
For ${d>1}$, it also includes condensation defects, which are constructed by gauging a higher-form symmetry on a lower-dimensional submanifold~\cite{Gaiotto:2019xmp, Roumpedakis:2022aik}.
The associated SymTFT is a ${(d+2)}$d TFT whose topological defects are described by the generalized Drinfeld center $\cZ(\cS)$, which is a braided fusion $d$-category.

Here we focus on symmetries in ${(2+1)}$d described by a fusion 2-category~\cite{DouglasReutter}.
The objects of $\cS$ describe symmetry surface defects, while the lines on the identity surface form a braided fusion category $\Omega\cS$ describing the 1-form symmetry.
For a finite $G$ 0-form symmetry with anomaly ${[\varpi]\in \cH^4(G,\Uone)}$, ${\cS=2\Vec_G^\varpi}$ and ${\Omega\cS\simeq\Vec}$.
When $G$ is anomaly-free, gauging $G$ gives rise to a $2\Rep(G)$ symmetry with ${\Rep(G)}$ 1-form symmetry.
Unlike fusion categories, fusion 2-categories are classified~\cite{Decoppet:2024htz}. A fusion 2-category $\cS$ is of \textit{all-boson (AB) type} when the M\"uger center $\cZ_2(\Omega\cS)$, consisting of lines that braid trivially with every line in $\Omega\cS$, is Tannakian. It is of \textit{emergent-fermion (EF) type} when ${\cZ_2(\Omega\cS)}$ is super-Tannakian but not Tannakian. 
AB-type fusion 2-categories are classified by
\begin{equation}\label{ABData}
    \cS\longleftrightarrow
    (\cB,H\leq G,\rho,\varpi,\mu),
\end{equation}
where $\cB$ is an MTC, ${H\leq G}$ are finite groups, $\rho$ is an action of $H$ on $\cB$ by braided autoequivalences, ${[\varpi]\in\cH^4(G,\Uone)}$, and $\mu$ specifies a homotopy between the anomaly of $\rho$ and $\varpi|_H$.

\begin{marginnote}[]
\entry{Tannakian}{A symmetric fusion category is Tannakian if it is braided equivalent to $\Rep(G)$ for some finite group $G$.
It is super-Tannakian if it is braided equivalent to $\Rep(G,z)$, where $z$ is a central element satisfying ${z^2=1}$ and acting as fermion parity.}
\end{marginnote}

The SymTFT $\fZ(\cS)$ is a ${(3+1)}$d TFT with a canonical topological boundary $\Bsym_\cS$ whose topological defects form $\cS$.
When $\cS$ is equipped with a spherical structure, it can be realized by the Douglas--Reutter state-sum construction, generalizing the Turaev--Viro construction discussed in Sec.~\ref{FusionCatSymTFT}~\cite{DouglasReutter}.
For ${\cS=2\Vec_G^\varpi}$, it gives $G$ Dijkgraaf--Witten theory with twist ${\varpi}$.
More generally, every AB-type fusion 2-category $\cS$ has ${\cZ(\cS)\simeq\cZ(2\Vec_G^\varpi)}$ ~\cite{Decoppet:2022}, so its SymTFT is equivalent to a Dijkgraaf--Witten theory (up to possible additional invertible TFTs)~\cite{Bhardwaj:2024qiv, Bullimore:2024khm, Xu:2024pwd, Bhardwaj:2025piv}.
This agrees with the classification of ${(3+1)}$d topological orders whose point excitations are all bosons~\cite{Lan:2018vjb}.
EF-type fusion 2-category symmetries instead have SymTFTs with emergent fermionic particles~\cite{Lan:2018bui}.
In the following, we restrict to AB-type fusion 2-category symmetries.

\noindent{\bf Topological Boundary Conditions.}
The Dirichlet boundary $\fB_{\Dir}$ of $\varpi$-twisted $G$ gauge theory, on which the Wilson lines of $\fZ(2\Vec_G^\varpi)$ can end, realizes a $2\Vec_G^\varpi$ symmetry. 
Every other gapped boundary can be obtained by stacking $\fB_{\Dir}$ with a (2+1)d TFT $\mathfrak{T}$ enriched by a ${H\leq G}$ symmetry whose anomaly cancels $\varpi|_H$, and then gauging the diagonal $H$ symmetry~\cite{Xu:2024pwd, Bhardwaj:2024qiv, Bullimore:2024khm, Wen:2025thg, Bhardwaj:2025piv}. The corresponding gapped boundary is
\be
\label{GappedBC3d}
\fB_{(H,\mathfrak{T})}=\frac{\fB_{\Dir}\boxtimes\mathfrak{T}}{H^{\rm diag}} \,.
\ee
The anyons of $\mathfrak{T}$ form an MTC $\cB$. 
The boundary $\fB_{(H,\mathfrak{T})}$ realizes the fusion 2-category symmetry with data~\eqref{ABData}, where $\rho$ describes the $H$ action and $\mu$ specifies the anomaly-cancellation data.
Every AB-type fusion 2-category symmetry can, therefore, be obtained from a 0-form symmetry by stacking with a TFT and gauging. 
When $\mathfrak{T}$ is an $H$-SPT, the boundary $\fB_{(H,\mathfrak{T})}$ is \textit{minimal}; this requires ${[\varpi|_H]=0}$.
Minimal boundaries can realize both 2-group and non-invertible symmetries.
When $\mathfrak{T}$ has intrinsic topological order, the boundary $\fB_{(H,\mathfrak{T})}$  is \textit{non-minimal}.
Since there are infinitely many $H$-enriched ${(2+1)}$d TFTs, there are infinitely many topological boundaries of a ${(3+1)}$d SymTFT~\cite{Bhardwaj:2024qiv}, unlike for the ${(2+1)}$d SymTFTs for fusion category symmetries.
Categorically, these boundaries are described by Lagrangian algebras in $\cZ(2\Vec_G^\varpi)$~\cite{Xu:2024pwd}. Those admitting a topological interface with $\fB_{\Dir}$  also admit a description in terms of module 2-categories over $2\Vec_G^\varpi$~\cite{Decoppet:2021skp, Decoppet:2023bay, Bhardwaj:2025jtf}.

\noindent{\bf Gapped Phases.}
Gapped phases with fusion 2-category symmetry $\cS$ are classified by the SymTFT sandwiches ${(\Bsym_\cS \mid \fZ(2\Vec_G^\varpi) \mid \fB_{(H,\mathfrak{T})})}$~\cite{Bhardwaj:2024qiv, Wen:2025thg, Bhardwaj:2025piv}.
Assuming the SymTFT is not invertible, its topological defect lines and surfaces that extend across the sandwich correspond to local and line order parameters, respectively, of the gapped phase~\cite{Bhardwaj:2023ayw, Bhardwaj:2025piv}.
For ${\cS=2\Vec_G^\varpi}$, the gapped phase $\fB_{(H,\mathfrak{T})}$ spontaneously breaks $G$ to $H$ with each vacuum realizing the $H$-enriched TFT $\mathfrak{T}$.
In this case, minimal physical boundaries correspond to SPT and SSB phases without topological order, whereas non-minimal physical boundaries describe gapped phases with intrinsic topological order.

\noindent{\bf Gapless Theories.}
Just as for ${(2+1)}$d SymTFTs, condensable algebras in $\cZ(\cS)$ organize the gapless physical boundaries of ${(3+1)}$d SymTFTs by defining a topological interface $\cI$ from $\fZ(\cS)$ to a reduced SymTFT specified by the local modules of this algebra~\cite{Decoppet:2023rlx, Kong:2024ykr}.
For the non-chiral interfaces studied in Ref.~\cite{Bhardwaj:2025jtf}, the reduced SymTFT is the Dijkgraaf--Witten theory $\fZ(2\Vec_{H/N}^\pi)$, where ${N\trianglelefteq H\leq G}$ and ${[\pi]\in \cH^4(H/N,\Uone)}$. 
Such an interface is specified by ${(H,N,\pi)}$ and a non-chiral $H$-symmetric TFT whose anomaly cancels the mismatch between the two bulk twists~\cite{Xu:2024pwd,Bhardwaj:2025jtf}. Placing an $H/N$-symmetric gapless theory with anomaly $[\pi]$ on the physical boundary of the reduced SymTFT completes the club sandwich, producing an $\cS$-symmetric gapless theory~\cite{Bhardwaj:2023bbf,Bhardwaj:2025jtf}. For minimal interfaces with ${\cS=2\Vec_G}$ and nontrivial $[\pi]$, these constructions yield igSPTs for ${H=G}$ and igSSBs for ${H\subsetneq G}$, respectively~\cite{Thorngren:2020wet,
Wen:2024qsg,Bhardwaj:2025jtf}.

\noindent{\bf Anyon Proliferation.}
The ${(3+1)}$d SymTFT has also been used to construct field theories for proliferation transitions of abelian and non-abelian anyons in ${(2+1)}$d topological orders~\cite{Schafer-Nameki:2026hgx,Cheng:2026hgu,Gai:2026vgz}.
Consider a proliferation transition between the ${(2+1)}$d topological orders $\cT$ and $\cT/\cA$, where $\cA$ is a condensable algebra in $\cT$.
The relevant SymTFT to study this transition is constructed from the fusion 2-category obtained by delooping and completing the braided fusion subcategory ${\langle\cA\rangle\subseteq\cT}$. The symmetry boundary is taken as its canonical gapped boundary, while the physical boundary is chosen so the SymTFT sandwich realizes $\cT$. 
With the bulk SymTFT and physical boundary held fixed, coupling matter fields to the topological lines on the symmetry boundary can drive a boundary phase transition whose topological limit yields a new SymTFT sandwich realizing $\cT/\cA$.
Determining whether this is a direct continuous transition and, if so, its universality class requires further analysis of the resulting field theory.

\noindent{\bf Even Higher Dimensions.}
Very little is known about the classification of general fusion $d$-categories. For odd $d$, there are non-group-theoretical symmetries with duality defects analogous to Tambara-Yamagami fusion categories. Explicit constructions for SymTFTs of non-invertible symmetries have appeared for ${(3+1)}$d~\cite{Kaidi:2022cpf,Antinucci:2023ezl,Cordova:2023bja} and ${(5+1)}$d symmetries~\cite{Lawrie:2023tdz,Apruzzi:2024cty}.

\section{Further Developments and Open Problems}\label{sec:outlook}

We conclude by surveying other extensions and applications of the SymTFT framework, together with open problems and directions for future research.

\noindent{\bf Microscopic constructions.}
The SymTFT also provides a systematic framework for constructing quantum lattice models with generalized symmetries.
In ${(1+1)}$d, the sandwich construction produces Hamiltonians with fusion category symmetries~\cite{Aasen:2020jwb,Bhardwaj:2024kvy}, which generalize earlier anyon-chain models~\cite{Feiguin:2006ydp}.
They have been used, for example, to construct Haagerup-symmetric critical anyon-chain models~\cite{Bottini:2025hri}.
Tensor-network methods provide explicit realizations of the SymTFT bulk--boundary correspondence~\cite{Bultinck:2015bot}.
Matrix product states with matrix product operator symmetries describe gapped boundary states~\cite{Garre-Rubio:2022uum}, while strange-correlator constructions realize critical boundary theories~\cite{Vanhove:2018wlb} and are used for numerical searches for new critical points~\cite{Hung:2025gcp}.
In ${(2+1)}$d, the sandwich construction leads to fusion-surface models with fusion 2-category symmetry~\cite{Inamura:2023qzl}.
These models have been used to construct commuting-projector Hamiltonians and tensor-network ground states for non-chiral gapped phases with AB-type fusion 2-category symmetries~\cite{Inamura:2025cum}.
Fusion-surface models also provide microscopic realizations of chiral topological order~\cite{Ueda:2024jzj, Eck:2024myo}.
It would be interesting to implement these models on quantum-simulation platforms, following proposals such as~\cite{Warman:2024lir}, and to further develop their numerical study to systematically search for new phases and critical points.

\noindent{\bf Continuous Symmetries.} 
The SymTFT of a continuous symmetry is necessarily a TFT with infinitely many simple topological defects, which goes beyond the UMTC framework in ${(2+1)}$d.
Developing the general categorical framework that takes its place is an ongoing effort~\cite{
MS250314596,JLTWZ250913170}. 
Nevertheless, explicit Lagrangian field theories of the SymTFT for compact connected Lie-group symmetries have been proposed~\cite{BS240106128,AB240110165,BDM240212347,ABR240801418,
GGE241115126,JLTWZ250304546, ADG250907965}.
These generalize the finite-group SymTFT construction by starting with the SPT encoding the symmetry's anomaly and then \textit{flat gauging} the symmetry.
The flatness constraint ${F_a=\dd a-\ii a\wdg a=0}$ makes Wilson loops topological defects, and the Dirichlet boundary condition of the resulting topological gauge theory is the symmetry boundary for the compact connected Lie-group symmetry.
For example, the SymTFT for a U(1) symmetry in a bosonic ${(1+1)}$d theory with 't~Hooft anomaly ${k\in \Z}$ is
\begin{equation}\label{UoneSymTFT}
    S_{\Symtft}^{\Uone}[M_3]
    =\frac{\ii}{2\pi}\int_{M_3}b\wdg\dd a + \frac{\ii k}{2\pi}\int_{M_3}a\wdg\dd a,
\end{equation}
where $a$ is a $\Uone$ gauge field and $b$ is an $\R$ gauge field.
The first term enforces the flatness condition ${\dd a = 0}$, while the second term encodes the U(1) symmetry's anomaly. For ${k=0}$, the topological defect lines are ${W_n(C)=\mathrm{exp}(\ii n\oint_C a)}$ and ${U_\theta(C')=\mathrm{exp}(\frac{\ii\theta}{2\pi}\oint_{C'}b)}$, which have mutual braiding statistics $\e^{\ii n \theta}$.
Compactness of $a$ quantizes ${n\in\Z}$ and identifies ${\theta\sim\theta+2\pi}$; the non-compactness of $b$ allows $\theta$ to vary continuously. 
The Dirichlet symmetry boundary ${a|_{\Bsym}=0}$ realizes $U_\theta$ as U(1) symmetry defects.
With suitable physical boundary conditions, the $\Uone$ SymTFT sandwich can realize compact boson CFTs~\cite{ABR240801418}. 
The SymTFT for both momentum and winding U(1) symmetries is a straightforward generalization of~\eqref{UoneSymTFT}, and captures flat gaugings that change the compact boson radius and corresponding non-invertible T-duality symmetries~\cite{ACGHP240911822}.
Higher-dimensional and higher-form symmetry generalizations similarly realize field theories for Goldstone bosons~\cite{ABR240801418,BDM250910343}, and can also capture the non-invertible $\mathbb{Q}/\Z$ chiral symmetry of ${(3+1)}$d QED-like theories~\cite{BS240106128,AB240110165}.

\noindent{\bf Fermionic systems.}
Finite symmetries of ${(1+1)}$d fermionic systems are described by fermionic fusion supercategories~\cite{Inamura:2022lun, Bhardwaj:2024ydc}. 
In the absence of a gravitational anomaly, they can be realized by a ${(2+1)}$d bosonic SymTFT with a fermionic symmetry boundary, which has local boundary fermions on which fermionic anyons can end~\cite{GK200805960, Wen:2024udn}.
(With a gravitational anomaly, the SymTFT can still be bosonic, but its symmetry boundary is then an interface to an invertible fermionic TFT~\cite{BoyleSmith:2024qgx}.)
For example, in the absence of a gravitational anomaly, the SymTFT of $\mathbb{Z}_2^F$ fermion number parity symmetry is $\Z_2$ gauge theory with an $f$-condensed symmetry boundary $\Bsym$. Replacing $\Bsym$ with the $e$-condensed boundary implements bosonization, while the reverse implements fermionization~\cite{Vanhove:2024lmj}. These symmetries can generally be obtained from bosonic fusion categories with anomaly-free $\mathbb{Z}_2$ subcategories via fermionization. This framework has been used to describe fermionic SPT phases, intrinsically gapless SPTs, and symmetry-enriched critical points~\cite{Wen:2024udn,Huang:2024ror,Bhardwaj:2024ydc}, and provides constructions of fermionic CFTs~\cite{Kawabata:2025hfd}.
Related ${(3+1)}$d SymTFT constructions have been used to describe ${(2+1)}$d superconducting phases and their Majorana-cone transitions~\cite{Ji:2024oky}.
One unexplored direction is finding applications of this extension of the SymTFT to phases and transitions of strongly correlated electronic systems, such as the Fermi--Hubbard model and non-Fermi liquids.

\begin{marginnote}[]
\entry{Fermionic fusion supercategory}{
A fusion supercategory is a generalization of a fusion category whose morphism spaces are supervector spaces, decomposed into even (bosonic) and odd (fermionic) parts.
A fermionic fusion supercategory has additional structure, including a distinguished invertible object representing the fermion-parity symmetry defect.
}
\end{marginnote}

\noindent{\bf Boundaries.}
The SymTFT framework also characterizes symmetry actions on boundary conditions and interfaces~\cite{Bhardwaj:2024igy, Choi:2024tri, GarciaEtxebarria:2024jfv}.
In the boundary SymTFT construction, an additional topological boundary $Q$ connecting $\Bsym_{\cS}$ and $\Bphys_{\cT}$ gives a boundary $B$ for the interval-compactified theory $\mathcal{T}$. Its junction with $\Bsym_{\cS}$ describes a boundary condition multiplet and its symmetry transformations, while its junction with $\Bphys_{\cT}$ specifies its physical realization.
This construction captures SPT edge modes and symmetry actions on boundaries of symmetry-breaking and gapless phases~\cite{Bhardwaj:2024igy}.
In ${(1+1)}$d, related constructions realize strip algebras, which organize solitons into symmetry multiplets~\cite{Cordova:2024iti}, as well as generalized tube algebras, which describe local boundary operators~\cite{Choi:2024tri}.
It would be interesting to apply these boundary SymTFT constructions to study how non-invertible symmetries constrain impurity screening and identify new boundary universality classes.

\begin{marginnote}[]
\entry{Boundary SymTFT}{
\begin{equation*}
\begin{tikzpicture}
  \begin{scope}[shift={(0,0)}]
    \fill[SymTFTcolor] (0,0) rectangle (1.25,2);
    \draw[very thick] (0,0)--(0,2)--(1.25,2)--(1.25,0);
    \node at (1.25/2,1) {$\fZ(\cS)$};
    \node[below] at (0.125,0) {$\Bsym_{\cS}$};
    \node[below] at (1.25,0) {$\Bphys_{\cT}$};
    \node[above] at (1.25/2,2) {$Q$};
    \filldraw[fill=white,draw=black]
      (0,2) ++(-0.05cm,-0.05cm) rectangle ++(0.1cm,0.1cm);
    \filldraw[fill=white,draw=black]
      (1.25,2) ++(-0.05cm,-0.05cm) rectangle ++(0.1cm,0.1cm);
  \end{scope}
  \begin{scope}[shift={(2,0)}]
    \node at (-0.4,1) {$=$};
    \draw[very thick] (0,0)--(0,2);
    \filldraw[fill=white,draw=black]
      (0,2) ++(-0.05cm,-0.05cm) rectangle ++(0.1cm,0.1cm);
    \node[above] at (0,2) {$B$};
    \node[below] at (0,0) {$\cT$};
  \end{scope}
\end{tikzpicture}
\end{equation*}
}
\end{marginnote}

\noindent{\bf Lower-form symmetries.}
Generalized symmetry also extends to negative form degrees.
For $(-1)$-form symmetries, scalar background fields are coupling parameters, such as $\theta$-angles, and symmetry defects are spacetime-filling insertions that can relate theories at different values of these coupling parameters~\cite{Cordova:2019jnf}.
They can be incorporated into the SymTFT, which has been used to describe their mixed anomalies, relation to decomposition, and non-invertible generalizations~\cite{Aloni:2024jpb, Najjar:2024vmm,
Lin:2025oml,Robbins:2025urk}.
It would be interesting to extend these applications to study phenomena in parameterized many-body Hamiltonians, such as higher Berry curvature and topological pumping.

\noindent{\bf Invertible SymTFT.}
A SymTFT can be invertible, with no intrinsic bulk topological order.
For example, a 1-form symmetry in ${(2+1)}$d described by an MTC has an invertible ${(3+1)}$d Crane--Yetter SymTFT, realized microscopically by a Walker--Wang model~\cite{Crane:1993if,Walker:2011mda}.
Invertible SymTFTs can nevertheless admit inequivalent topological boundaries supporting different topological orders, e.g., via~\eqref{GappedBC3d} for ${(3+1)}$d.
Unlike in topologically ordered SymTFTs, topological defects on $\Bsym$ of invertible SymTFTs do not need to stretch across the sandwich and end on $\Bphys$; instead, they can end in the bulk~\cite{ABB240406601}.
This implies that $\Bsym$ and $\Bphys$ factorize, so every symmetric theory has a factorized non-invertible TFT corresponding to $\Bsym$ that spontaneously breaks the symmetry. Thus, nontrivial symmetries with invertible SymTFTs are anomalous, and the anomaly enforces spontaneous symmetry breaking.
Further consequences of this boundary structure of the invertible SymTFT deserve further investigation, and it would be interesting to understand whether it can be used to constrain gapless theories and their phase transitions.

\noindent{\bf Mixed state symmetries.}
For a unitary symmetry operator $U$, a density matrix $\rho$ has $U$ as a weak symmetry if ${U\rho\, U^\dag = \rho}$, and $U$ as a strong symmetry if ${U\rho\propto \rho}$.
Mixed states can realize strong-to-weak SSB and SPT orders involving combinations of strong and weak symmetries that are not found in pure states~\cite{Ma:2024kma,Lessa:2024gcw}. 
These notions have now been extended to non-invertible symmetries: every mixed state ${\rho=\sum_ip_i|\psi_i\rangle\langle\psi_i|}$ with strong fusion-category symmetry $\cS$ has a corresponding vector ${|\rho\rangle\rangle \propto \sum_ip_i|\psi_i\rangle|\psi_i^*\rangle}$ in a doubled Hilbert space with symmetry ${\cS\boxtimes \overline{\cS}}$, where $\overline{\cS}$ denotes the complex-conjugate fusion category. Not every vector in the doubled Hilbert space corresponds to a Hermitian, positive-semidefinite mixed state.
The ${(1+1)}$d SSB and SPT phases of mixed states with fusion category symmetry can be classified using the SymTFT of ${\cS\boxtimes \overline{\cS}}$~\cite{Qi:2025tal, Schafer-Nameki:2025fiy, Luo:2025phx}. 
This proposed classification is based on the Lagrangian algebras of the doubled SymTFT $\cZ(\cS\boxtimes \overline{\cS})$ subject to additional constraints from Hermiticity and positive-semidefiniteness. 
It would be interesting to construct parent Lindbladians and quantum channels that prepare the corresponding phases.

\noindent{\bf Spacetime symmetry.}
One approach to incorporating spacetime symmetries into the SymTFT framework is to enrich the SymTFT of an internal symmetry by a spacetime symmetry~\cite{PAL250702036,BJ260326884,WS260328720}. The enrichment data encodes interplays between the spacetime and internal symmetries, such as Lieb--Schultz--Mattis anomalies~\cite{Pace:2024acq,PAL250702036}.
For lattice translation symmetry, this enrichment can cause the SymTFT to be a foliated field theory, which is topological along the interval direction of the sandwich construction but non-topological in other directions.
Similar foliation structures appear in SymTFTs for subsystem symmetries~\cite{CJ231001474,ABM250411449,JJ250522261, BM260203926}, illustrating how the SymTFT sandwich construction requires topological invariance only along the interval direction.
An alternative approach for \emph{continuous} spacetime symmetries uses BF theories for the spacetime symmetry group with additional Chern--Simons terms encoding conformal and gravitational anomalies~\cite{ADG250907965}.
In certain dimensions, these theories are related to topological limits of gravity. 
A more systematic treatment of lattice symmetries, including subsystem symmetries, is an interesting direction for future work. 
Furthermore, while SymTFT constructions have been developed for the X-cube model and Haah's cubic code~\cite{Hu:2026dvc}, it remains an open question whether every fracton model with a gapped boundary can be a SymTFT for an exotic lattice symmetry.

\subsection*{Disclosure Statement}
The authors are not aware of any affiliations, memberships, funding, or financial holdings that
might be perceived as affecting the objectivity of this review.

\subsection*{Acknowledgments}

We are deeply grateful to 
{\"O}mer Aksoy,
Andrea Antinucci, 
Fabio Apruzzi, 
Ibou Bah, 
Lakshya Bhardwaj, 
Lea Bottini, 
Federico Bonetti, 
Xie Chen,
Christian Copetti, 
Clay Córdova,
Arkya Chatterjee,
Thibault D\'ecoppet, 
Yuhan Gai, 
I{\~n}aki Garc{\'\i}a Etxebarria, 
Saghar Hosseini, 
Sheng-Jie Huang, 
Kansei Inamura, 
Wenjie Ji,
Ho Tat Lam,
Michael Levin, 
Daniel Pajer, 
Shu-Heng Shao,
Apoorv Tiwari, 
Alison Warman, 
Rui Wen, 
Xiao-Gang Wen,
and Matthew Yu
for the many stimulating conversations and collaborations that have shaped our understanding of the SymTFT.
S.D.P. is supported by the U.S. Department of Energy and the Marvin L. Goldberger Membership, William Loughlin Membership, and IBM Einstein Fellowship Fund at the Institute for Advanced Study. 
S.S.-N. is supported by the UKRI Frontier Research Grant, underwriting the ERC Advanced Grant ``Generalized Symmetries in Quantum Field Theory and Quantum Gravity.'' 
C.Z. is supported by NSERC grants RGPIN-2026-07143 and DGECR-2026-00535. 
We acknowledge use of ChatGPT (GPT-5 and later models) and Claude (Opus 5.5) to improve the writing, assist with literature searches, and audit the manuscript; we take full responsibility for the content, accuracy, and integrity
of the review.

\bibliographystyle{ytphys}

\bibliography{refs.bib}

\end{document}